\documentclass[aip,jcp,reprint]{revtex4-1}

\usepackage{amsmath, amsfonts,graphicx,color}

\begin{document}

\newcommand{\todo}[1] {{\color{black} #1}}
 
\newcommand{\LRC}{$L \leftrightarrow R$}
\newcommand{\I}{\mathcal{I}}
\newcommand{\E}{\mathcal{E}}
\newcommand{\C}{\mathcal{C}}
\newcommand{\OO}{\mathcal{O}}
\newcommand{\R}{\mathcal{R}}
\newcommand{\SC}{\mathcal{S}}

\title{A Spin--1 Representation for Dual--Funnel Energy Landscapes }

\author{Justin E. Elenewski}

\affiliation{Center for Nanoscale Science and Technology, National Institute of Standards and Technology, Gaithersburg, MD 20899, USA}

\affiliation{Maryland Nanocenter, University of Maryland, College Park, MD 20742, USA}
  
\author{Kirill A. Velizhanin}

\affiliation{Theoretical Division, Los Alamos National Laboratory, Los Alamos, NM 87545 USA}

\author{Michael Zwolak}

\email{mpz@nist.gov}

\affiliation{Center for Nanoscale Science and Technology, National Institute of Standards and Technology, Gaithersburg, MD 20899, USA}

\begin{abstract}
The interconversion between left-- and right--handed helical folds of a polypeptide defines a dual--funneled free energy landscape.  In this context, the  funnel minima are connected through a continuum of unfolded conformations, evocative of the classical helix--coil transition.  Physical intuition and recent conjectures suggest that this landscape can be mapped by assigning a left-- or right--handed helical state to each residue.  We explore this possibility using all--atom replica exchange molecular dynamics and an Ising--like model, demonstrating that the energy landscape architecture is at odds with a two--state picture. A three--state model -- left, right, and unstructured -- can account for most key intermediates during chiral interconversion. Competing folds and excited conformational states still impose limitations on the scope of this approach. However, the improvement is stark: Moving from a two-state to a three-state model decreases the fit error from 1.6 $k_B T$ to 0.3 $k_B T$ along the left-to-right interconversion pathway. 
\end{abstract}
\maketitle

\section{Introduction}

\par
\begin{figure}[b] 
\includegraphics[width=0.8\columnwidth]{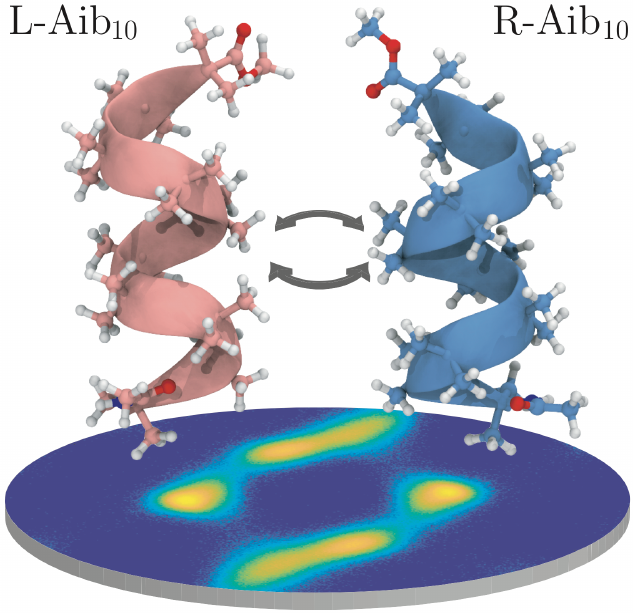}
\caption{ Interconversion between left-- and right--handed helical conformations in an Aib$_{10}$ peptide foldamer.} 
\label{fig:galaxy}
\end{figure}

 While most biomolecular helices possess a right--handed orientation, a small minority are predisposed to assume a left--handed fold.\cite{Novotny2005}  This chiral propensity may be enhanced through the introduction of achiral amino acids, such as 2--aminoisobutyric acid (Aib) ---  noted for its ability to induce a prominent left--handed helical population in biomimetic peptides (Fig. \ref{fig:galaxy}).\cite{Venkatraman2001}  While useful as a probe of biological organization, this observation also constitutes a general rule: a helix--forming polymer will demonstrate a preferred axial chirality when it is constructed from chiral blocks.  In contrast, polymers derived from achiral blocks exhibit a degeneracy of  left-- ($L$) and right--handed ($R$) helical folds, forming an effective two--state system.\cite{Hummel1987}  Deviations from this behavior must be associated with induced chirality,  either from the solvent environment or from chiral structural elements flanking the polymer.

\par This tunable two--state character has been exploited in the design of foldamer--based nanodevices.\cite{Hill2001, Martinek2012, LeBailly2016}  By introducing structural features that extend beyond standard biological motifs, these materials can be engineered to include distinctive photoreactive,\cite{DePoli2016} thermoresponsive,\cite{Sakai2006, Li2014} pore--forming,\cite{Jones2015} and ligand binding functionalities.\cite{Suk2011, Horeau2017}  Aib--containing decamers, in particular, form stable helices and have found use as actuators within biomimetic receptors \cite{Brioche2015, LeBailly2016, Lister2017} and photoswitches.\cite{DePoli2016} In this case, triggering a conformational shift in the N--terminal sensor domain --- through either ligand binding or photoexcitation ---  \todo{presumably biases the free energy landscape toward the opposite helical chirality, analogous to solvent--driven transitions in polyproline peptides.$\cite{Moradi2009, DePoli2013}$ This shift results in an $L \leftrightarrow R$ interconversion, the subsequent} structural reorganization of a covalently linked C--terminal reporter and the \todo{ultimate} detection of an activation signal.  While promising as biomimetic devices, these systems also afford a fundamental opportunity to explore the translation of local molecular interactions into folding and function, outside the confines of natural biological systems.\cite{Goodman2007}

\todo{\par The static and dynamic properties of a macromolecule are dictated by the architecture of its potential energy landscape.\cite{Bryngelson1987,Bryngelson1989,Leopold1992}  In the case of biomolecules, these landscapes are minimally frustrated, containing the smallest possible set of competing low--energy conformations, separated by high potential barriers.  This principle often affords a sharply funneled profile, with the so--called native state lying at the global minimum and successive `excited' states occupying higher--lying local minima.  These minima collectively define the stable conformers of the molecule, while barriers in the landscape dictate the kinetic profile for conformational transitions. 

\par Multifunctional biomolecules deviate from this paradigm, delivering a potential energy landscape that contains separate funnels for each functional conformer.\cite{Roder2018}  This organizational principle extends to helical foldamers, where left-- and right--handed  orientations correspond to separate funnel minima in a dual--funnel energy landscape.\cite{Chakrabarti2011, Olesen2013}  While increasingly recognized in a biological context, the initial theoretical foundation for these systems was derived to describe dual--funnel, solid--solid transitions of small Lennard--Jones (LJ) clusters.\cite{Wales1998, Doye1999}  These investigations delivered fundamental insight into the hallmarks of multifunnel landscapes, and illustrated how landscape features can mediate processes that range from kinetic trapping to phase changes in extremely finite systems.  Explorations of cluster landscapes have likewise inspired, and serve as a benchmark for, ergodic sampling methods in molecular simulations,\cite{Calvo2000, Neirotti2000, Mandelshtam2006, Wales2006, Sharapov2007, Calvo2010, Wales2013, Sehgal2014, Wales2015} global optimization schemes,\cite{Oakley2013} and path--sampling frameworks for rare--event dynamics.\cite{Wales2002, Wales2004,Adjanor2006, Picciani2011,Cameron2014}  These computational advances have, in turn, facilitated more recent  studies of biological and biomimetic systems, including those undergoing helix chirality inversions\cite{Moradi2009, Chakrabarti2011, Olesen2013, Chakraborty2017} and containing bistable switching motifs.\cite{Kouza2012, Roder2017, Chakraborty2018}

}

\par  \todo{While certain biomolecules have been well--explored, a systematic, theory--guided approach to foldamer design is impeded by the complex nature of multifunnel energy landscapes.  In principle, these may only be mapped using costly numerical simulations.\cite{Hill2001, Doig2008}}  Nonetheless, a suitable analytical model --- parameterized using discrete calculations at the single--block level --- might dramatically accelerate these efforts while retaining  acceptable quantitative accuracy.  Numerous computational techniques have been applied to Aib--containing helices in order to understand the form that such a model should take.\cite{Paterson1981,Zhang1994, Burgi2001, Improta2001, Mahadevan2001, SchweitzerStenner2007, Grubisic2013, Grubisic2016}  Nonetheless, the majority of these are limited in their sampling of conformational dynamics during the $L \leftrightarrow R$ transition.   Motivated by studies of energy transport in the Aib$_9$ peptide foldamer,\cite{Botan2007, Backus2008, Backus2008b, Backus2009b,  Nguyen2010} more recent efforts have long--timescale molecular dynamics simulations, Markov state modeling, and principal component analysis (PCA) to these systems.\cite{Buchenberg2015,Sittel2017}  The resulting data suggest that Aib$_9$ has a conformational landscape where long timescale (ns to $\mu$s)  dynamics are ultimately slaved to short timescale (ps) hydrogen bonding transitions through a \todo {hierarchy of timescales --- a feature characteristic of systems with a multi--funnel architecture.\cite{Chan1993, Chan1994, Doye1999b}} A coarse--grained energy landscape was proposed, parameterized in terms of  backbone dihedral angles $(\phi,\psi)$, with conformational substates determined by discrete, per--residue changes in helical chirality.  In this manner, each residue is assigned to either a left ($L$) or right ($R$) conformational substate, though the authors did not make this statement quantitative.\cite{Buchenberg2015,Sittel2017}  \todo{While more elaborate coarse--grained models can capture  the  dynamics of helix chirality inversion,\cite{Olesen2013} it is unclear if this minimal description can do the same.}

\par To explore this physically intuitive proposal, \todo{and assess the limits of coarse--graining}, we have constructed an explicit representation of the $L/R$--model using an Ising--like Hamiltonian. Large--scale replica--exchange molecular dynamics (REMD) simulations and  an unsupervised clustering approach were employed for parameterization, affording a granular picture of the potential energy landscape.  Our observations indicate that a two--state representation is  insufficient to describe the structural diversity inherent in the \LRC\, transition.  A three--state, spin--1 model yields better performance by explicitly including unstructured regions. However, the model still exhibits systematic deviations from all--atom simulations.  These inconsistencies arise from numerous factors, including  the presence of `excited' conformational substates and the existence of distinct, competing helical folds.  Taken together, our observations provide a minimal bound on the complexity of analytical models that accurately describe simple foldamers such as Aib$_N$, where $N$ is the number of repeats, and attest to the necessity of explicit simulation in characterizing these systems.

\section{Theory and Methods}

\subsection{Spin Models}

\noindent  We consider an unbranched foldamer containing $N$ linearly--ordered blocks. In the simplest case, each of these blocks might be classified as either a left--handed ($L$) or right--handed ($R$) configuration according to its backbone dihedral angles (Appendix I).   An energetic gain of $J < 0$ per site is expected when consecutive blocks share the same helical orientation ($\dots RR \dots$ or $\dots LL \dots$) --- promoting the formation of homochiral domains  --- and an energetic penalty of $-J > 0$ encountered at a domain wall between left-- and right--helical regions ($\dots LR \dots$ or $\dots RL \dots$).  The simplest Hamiltonian that describes these interactions is a \emph{classical} `ferromagnetic' spin--1/2 Ising model,

\begin{equation} \label{spinham}
E_\mathcal{I} = \sum_{\alpha = 1}^{N-1} \, \hat{J} (\sigma_{\alpha + 1}, \sigma_\alpha)
\end{equation}

\noindent where the  spin $\sigma_\alpha \in \{L,R\}$ residing on block $\alpha$ is assigned to either a right--handed $(R)$ or left--handed $(L)$ configuration.  The spin-spin coupling matrix $\hat{J}$ is symmetric, and all elements are of the same magnitude: $J=\hat{J}(R,R)=\hat{J}(L,L) = -\hat{J}(L,R)=-\hat{J}(R,L).$ The supplementary material (SM) shows the effect of asymmetric couplings and discusses our approach to fitting  (SM Section IA, Figs. S1-S5). We will discuss the difficulties encountered with this simple approach.

\par The energy, Eq.~\eqref{spinham}, can be extended to a spin--1 scenario, where an unstructured random coil state $U$ exists alongside the $L$ and $R$ configurations ($\sigma_\alpha \in \{L,U,R\}$). We take the coil state to always refer to the unstructured polymer or region. For simplicity, we assume that extended coil stretches have no inter--site coupling $\hat{J}(U,U) = 0$, nor do they make an energetic contribution when contacting helical regions $\hat{J}(L/R,U) = 0$ (SM Section IB, Fig. S6, S7)   This construction differs from Zimm--Bragg\cite{Zimm1959} and Lifson--Roig\cite{Lifson1961} models for the helix--coil transition, where an additional statistical weight for nucleation is assigned to helix termini that flank unfolded regions.  The inclusion of a nucleation penalty $\hat{J}(L/R,U) \neq 0$, or a correction for flexible helix termini, has only a modest impact on our model (see the SM, Figs. S7-S8).

\par The persistence of extended coils will depend on a variety of factors, including the interaction of side chain and backbone atoms with the encapsulating solvent.  This effect may be captured through an on--site solvation energy $K_S$ that is associated with unstructured peptide regions.  We may also define a contribution $\hat{K}(L/U/R)$ that reflects the tendency of a given conformation to reside in rotameric minima (not including cooperative factors, such as hydrogen bonding), leading to an on--site Hamiltonian term:

\begin{equation}
E_\OO = \sum_{\alpha=1}^N \hat{K} (\sigma_\alpha).
\end{equation}

\noindent In practice, it is convenient to set $\hat{K}(L) = \hat{K}(R)  = 0$ and introduce a single nonzero on--site parameter $\hat{K}(U) = K_S - K_0$ that reflects the impact of solvation (and other factors that impact the ``on site'' energy of a block) on the extended coil state (see the SM, Figs. S9-S10).  Under these considerations $\hat{K}(U) < 0$ promotes and $\hat{K}(U) > 0$ penalizes the formation of extended coil regions.  For Aib peptides in chloroform, one would expect a $\hat{K}(U) > 0$ as contacts between the polar backbone amides and the nonpolar solvent would be disfavored.

\par In a realistic foldamer, coiled regions will admit a multitude of conformations, while the comparatively rigid helical residues are likely to cluster around a single conformational state.  This behavior may be accommodated by introducing the contribution of rotameric entropy to the free energy of the system

\begin{equation} \label{entropy}
S_\R = k_B \sum_{i \in \C} (n_i - 1) \cdot s,
\end{equation}

\noindent where $n_i$ is the length of the $i$--th coiled stretch, $k_B \cdot s$ is a unit of conformational (rotameric) entropy, and $k_B$ is the Boltzmann constant.  The summation runs over indices in  the set of coiled regions, $\mathcal{C}$, defined as a repeat of two or more consecutive $U$ sites in a given peptide conformation.  The high flexibility of helix termini, coupled with their ambiguous dihedral assignments, necessitates their exclusion when determining membership in $\mathcal{C}$.   The appearance of  $(n_i - 1)$ in Eq. (\ref{entropy}) and the definition of $\mathcal{C}$ both require an explanation.  While a single $U$ site results in a kinked helix -- and a slight increase in entropy  -- broad conformational (rotameric) sampling only occurs when there is junction between two $U$ sites \todo{(an upper bound for the kink entropy is given by Fig. S11b; however the minute conformational variability observed in all--atom MD is beyond the scope of our model}).  As a consequence of this, an expression scaling as $n_i$ results in an overestimation of the entropy for otherwise structured states, leading to the use of an $(n_i - 1)$ term (see the SM, Fig. S11).

\par These considerations collectively define the relevant contributions to the Gibbs free energy for a solvated, helical foldamer, containing both helical and unfolded segments:

\begin{equation} \label{fullmodel}
G_C = E_\OO + E_\I  - TS_\R + pV.
\end{equation}

\noindent As a matter of convention, the free energy of the $j$--th spin configuration $\Delta G_{C,j} = G_{C,j} - G_{C,0}$ will be measured relative to the absolute free energy $G_{C,0}$ of the lowest energy spin configuration(s)  in a given ensemble.  The Hamiltonian components of this model are exactly solvable, and the partition function may be evaluated using transfer matrix techniques.  Equation (\ref{fullmodel}) includes a contribution from volume, $pV$, where $p$ ($V$) is the pressure (volume), that is necessary for completeness.  \todo {We may alternatively drop this explicit dependence,  effectively wrapping this parameter into other terms appearing in Eq. (\ref{fullmodel}) by fitting simulation data.  This will be addressed later.}

\subsection{All--Atom Simulations}

\par Molecular dynamics simulations were performed using a ten residue stretch (Aib$_{10}$) of 2--aminoisobutyric acid, with initial backbone dihedrals assigned from the Dunbrack rotamer library for a right--handed $3_{10}$ helix.\cite{Shapovalov2011}  The model was embedded in a (5 nm)$^3$ cubic cell, containing 922 chloroform molecules, and packed to  the density of bulk solvent.\cite{Martinez2009}  C-- and N--termini were methylated and acetylated, respectively, and simulation physics described using the CHARMM36 force field\cite{Best2012} and the LAMMPS package.\cite{Plimpton1995}  CHARMM cross--term map (CMAP) corrections were not employed.\cite{MacKerell2004}  While this modification has been demonstrated to improve $\alpha$--helix folding cooperativity,\cite{Freedberg2004, Best2012} it dramatically overestimates $\alpha$--helical character for model helices.\cite{Patapati2011} Since Aib$_{10}$ is characterized by nontrivial $3_{10}$--helical content, this would have  questionable transferability  without major reparameterization.

\par Lennard--Jones and Coulomb interactions were computed using conventional CHARMM pair potentials, conjoined with an additional electrostatic damping term to maintain compatibility with particle--particle--particle mesh summation (force cutoff = $6.95 \times 10^{-3}$ pN).  Switching functions were employed to rescale coupling between atomic pairs separated by more than 1.0 nm, with the interatomic potential vanishing beyond 1.35 nm separation.  A timestep of 1.0 fs was used for all calculations within a scheme\cite{Shinoda2004} that employs a velocity Verlet integrator,  modified Nos\'{e}--Hoover thermostat and Martyna--Tobias--Klein barostat,\cite{Martyna1994} alongside a Parrinello--Rahman representation\cite{Parrinello1981} for the strain energy, allowing us to reproduce the correct probability distribution for the isobaric--isothermal (NPT) ensemble.  The damping period of the thermostat was set to 100 fs, while the damping period of the barostat was set to 1000 fs and coupled to a chain of eight members.   Isotropic cell fluctuations were allowed in all directions, and initial velocities assigned according to Gaussian distributions for both linear and angular momenta at a given temperature.

 \par The Aib$_{10}$ model was subjected to an initial 22 ns NPT equilibration, providing a starting point for subsequent replica exchange  simulations.  REMD runs were performed using 48 replicas in an NPT scheme,\cite{Okabe2001, Mori2010} with temperatures spanning between $T_i = 230$~K and $T_f = 465$~K.  Exchanges were attempted every 500~fs, yielding an acceptance ratio of 32 \% for all simulations.  REMD simulations were equilibrated for a further 150~ns, followed by a 500~ns production run.  Conformational clusters were determined using a running $k$--means scheme and a metric based on the root-mean-square deviation (RMSD) of heavy backbone atoms (0.26~nm, cutoff radius).  \todo{This procedure is sufficient to converge the energies of ground--state clusters --- presumed to be degenerate at all temperatures --- to within a deviation of 0.07~$k_B T$ when sampling below 300~K.  The persistence of this criterion for over 50 ns of sampling was taken as a hallmark for convergence, as force field deficiencies may prevent complete degeneracy from occurring on an accessible timescale during our simulations.\cite{Buchenberg2015}}

\section{Results and Discussion}

\subsection{Aib$_{10}$ Energy Landscape}

\par The conformational dynamics of Aib--based polypeptides have been experimentally characterized in a spectrum of solvents. However, their behavior in chloroform has a particularly distinguished history.  Under these conditions, it has been proposed that a glass--like dynamical transition occurs in Aib-Ala-(Aib)$_6$ derivatives,\cite{Botan2007,Backus2008,Backus2008b,Backus2009,Backus2009b} as reflected through temperature--dependent energy transport, though the nature of this transition -- and the changes in energy transport -- remains contentious, even among the same group of authors.\cite{Nguyen2010,Kobus2010,Kobus2011,Buchenberg2015} More concretely, a nonpolar environment mimics the artificial membranes in which many foldamer--based molecular sensors are intended to operate.\cite{DePoli2016, Lister2017}  These factors -- combined with the \todo{absence of solvent hydrogen bonding} or complex electrostatics -- motivated us to adopt this model system.

\begin{figure}[t]
\includegraphics[]{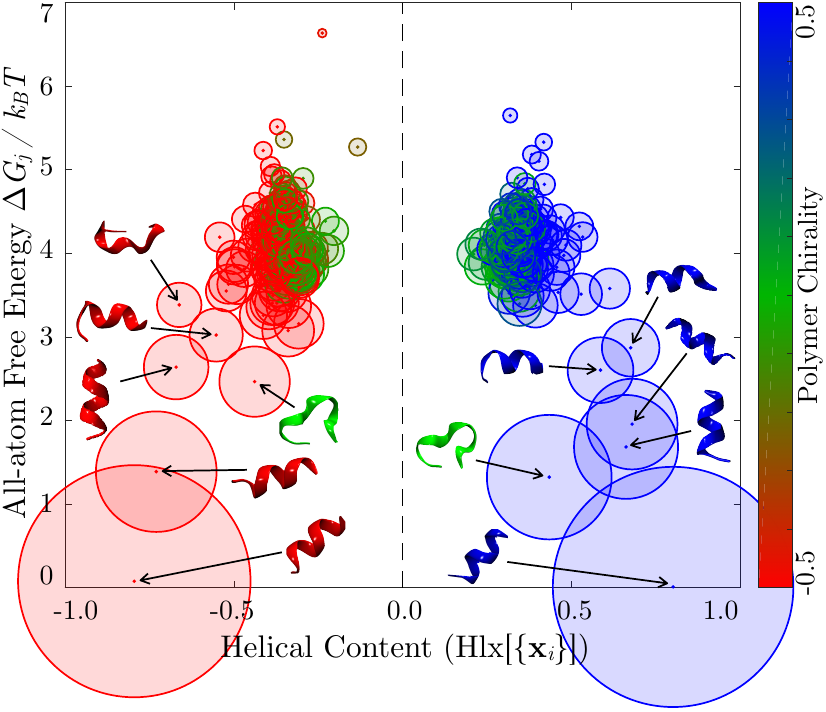}
\caption{Free energy landscape of Aib$_{10}$ at $T = 300$~K.  Each data point represents one of $N = 433 $ clusters, and the size of each point is scaled relative to the number of states contained within the respective cluster.} 
\label{fig:landscape}
\end{figure}

\par An enhanced--sampling approach was employed to facilitate the exploration of conformational space, using  all--atom REMD simulations to sample the NPT ensemble for an explicitly--solvated Aib$_{10}$ peptide.  This protocol generates approximately $10^6$ conformers for each temperature, however, the construction of an energy landscape requires either projection onto a lower--dimensional subspace or the reductive classification of these peptide conformations into a smaller set of states.  We adopted the latter approach --- a granular quantification of minor intermediates along the transition pathway.  \todo{While a full--dimensional approach might involve the use of disconnectivity graphs and related classification methods, a simple clustering scheme is sufficient for our purposes.\cite{Wales1998}} This character may be obscured when using PCA--based methods.  Classification was accomplished using  a $k$--means clustering scheme, in which a given trajectory frame is assigned to a cluster only if the running intra--cluster RMSD is less than 0.26~nm from the cluster centroid.  For a 500~ns REMD trajectory, this affords clusters containing $\approx 400$ states at each temperature that is sampled between 230~K and 330~K.  With this classification in hand, a relative free energy 

\begin{equation}\label{energyProb}
\Delta G_{jk} = -k_BT \log\, [P_k / P_j]
\end{equation}

\noindent may be calculated between states of populations $P_j$ and $P_k$, respectively.  For simplicity, we assume that free energies are measured with respect to the most populous cluster in each ensemble, allowing $\Delta G_j$ to be indexed by a single parameter.

\par REMD simulations reveal an energy landscape that contains two distinct folding funnels, corresponding to left-- and right--handed conformations of the peptide helix (Fig. \ref{fig:landscape}).  The native helical states in each funnel are nearly isoenergetic, however, a degree of asymmetry exists in distribution of clusters.  This is likely a collective effect, resulting from clustering artifacts, initial conditions, sampling limitations, and a degree of bias induced by the terminal patches that maintain electrostatic neutrality.  Interestingly, (a more sizable) asymmetry was  observed in a previously reported map of the Aib landscape derived following  long--timescale MD simulations and dimensional reduction of the data.\cite{Buchenberg2015}  On a separate note, the gap in helical content between left-- and right--handed funnels is intrinsic to the function $\text{Hlx}[\{\mathbf{x}_i\}]$ (see appendix), which accounts for oriented twists in the peptide backbone, reflecting an `intrinsic helical content' associated with gyration of the peptide chain.

\par  While the precise cluster assignments in each funnel are not identical, the overall distribution of these states remains quite similar.  Both funnels are dominated by highly--populated clusters of helix--rich states at low energies, expanding into a large set of high--entropy states with a low helical content at higher energies.  This high energy region also contains `excited' helical states, in which an energetically unfavorable conformation is assumed while preserving the overall helical fold.  An apparent crossing between funnels, where the fold becomes largely unstructured, occurs around $T = 4\, k_BT$ consistent with prior PCA--based landscapes. \cite{Buchenberg2015, Sittel2017}  \todo{This system is distinguished from other foldamers by the achiral nature of Aib, affording a highly symmetric and degenerate energy landscape.  In contrast, helices that are derived from chiral blocks\cite{Moradi2009,Chakrabarti2011,Olesen2013} often demonstrate some degree of energetic asymmetry, while clusters such as LJ$_{38}$ possess energetically asymmetric basins with markedly different topographies.\cite{Doye1999}}

\begin{figure}[t]
\includegraphics[width=\columnwidth]{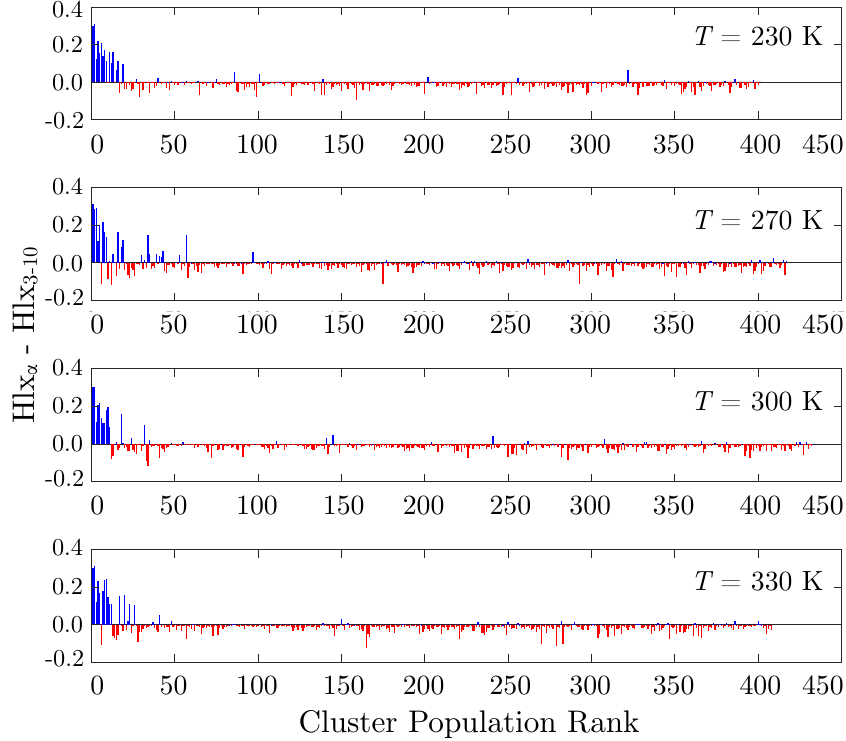}
\caption{Temperature--dependent helical content within the Aib$_{10}$ landscape, demonstrating an excess of either $\alpha$--helical (blue) or $3_{10}$--helical (red) character.  Clusters are ranked in order of decreasing population (i.e., increasing free energy).} 
\label{fig:helixContent}
\end{figure}

\begin{figure}[t]
\vspace{1.5\baselineskip}
\includegraphics[width=\columnwidth]{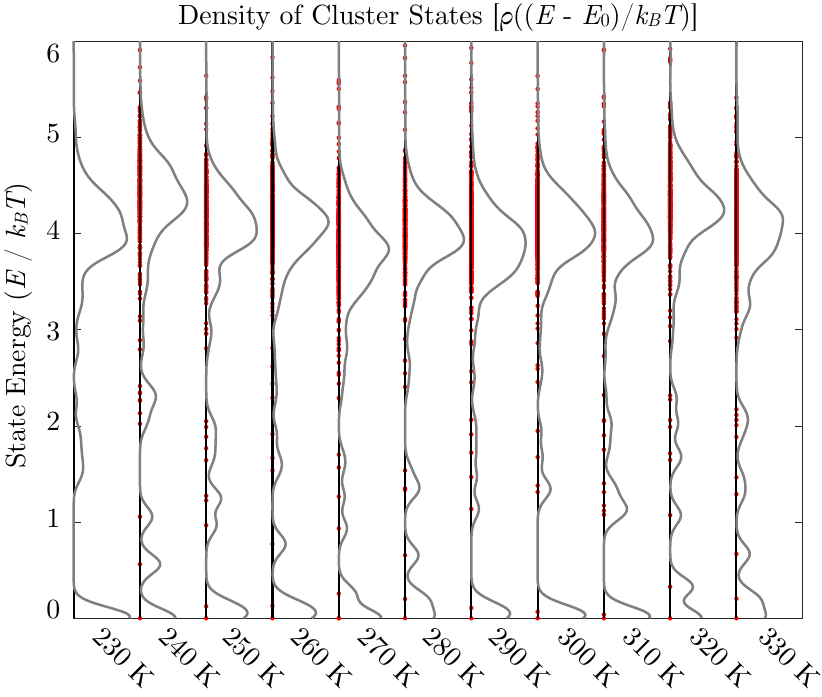}
\caption{Density of conformational clusters, populated according to the number of members in each cluster, for all--atom replica exchange simulations with base temperatures spanning between 220 K and 330 K.  The density of states was calculated using $\sigma^2 = 0.01$ as a smearing  parameter.  } 
\label{fig:dos}
\vspace{1.5\baselineskip}

\end{figure}

\par The low energy clusters in this ensemble are primarily $\alpha$--helical, transitioning to a combination of unfolded and $3_{10}$--character with increasing energy (Fig 3).  At first glance, this appears to contradict experiment, as crystallography, \cite{Francis1983, Bavoso1986, Gessmann2003} optical,\cite{Yasui1986a, Yasui1986b, Maekawa2008} and magnetic resonance spectroscopies suggest that short ($\leq 7$ residues) Aib peptides exist as $3_{10}$--helices in  weakly dielectric environments.  This picture is nuanced, as both $\alpha$-- and $3_{10}$--conformations -- with a dominant $\alpha$--helical population -- have been observed in polar and moderately polar solvents (water, DMSO).\cite{Bellanda2001, Burgi2001,Kuster2015, Gord2016}  Taken together, these data indicate that Aib conformations exhibit a high degree of environmental sensitivity, with $3_{10}$--helical content predominating for short helices in nonpolar media and at higher temperatures.\cite{Carlotto2007}   Since our ten--residue model is longer than the peptides employed in most experimental efforts, it is plausible that a greater degree of $\alpha$--helical content may occur in this system, with the early stages of folding templated by a $3_{10}$--helix.\cite{Topol2001} This behavior is consistent with MD simulations  using a custom, AMBER--based force field for Aib\cite{Grubisic2013,Grubisic2016} and long--timescale trajectories of Aib$_9$ dynamics using GROMOS96, which reveal an increasing proportion of $\alpha$--helical character for longer chains.\cite{Buchenberg2015} Nonetheless, the dominant helical fold, and any force--field dependence, will not markedly alter our conclusions. They key observation herein is the coexistence of two distinct helical populations with different physical characteristics.

\par The state distribution within the Aib$_{10}$ landscape is readily analyzed using the density of conformational cluster states $\rho(E)$ that are observed at each replica temperature

\begin{equation}
\rho(E) = \sum_{j=1}^{N_\text{C}} P_j \, e^{-(E-\Delta G_j)^2 /2\sigma^2}
\end{equation}

\noindent where $N_C$ is the number of $k$--means clusters calculated for a given replica, $P_j$ is the population of the $j$--th cluster, $\Delta G_j$ is the free energy of the $j$--th cluster,  and $\sigma^2$ accounts for the inter--cluster conformational variance (Fig. \ref{fig:dos}).  These data exhibit a weak temperature dependence, characterized by a 5 \% increase in the number of conformers lying below 4 $k_B T$  as the temperature is increased from 230 K to 330 K.  Notably, these data do not demonstrate any obvious signatures of a transition between 250 K and 270 K, where prior experiments and simulations have suggested at the Aib might undergo a protein dynamical transition.\cite{Botan2007,Backus2008,Backus2008b,Nguyen2010,Kobus2010,Kobus2011,Buchenberg2015} A dynamical transition cannot be excluded without a more comprehensive analysis, as these inherently coarse--grained clusters may obscure subtle changes associated with this phenomenon.   \todo{Furthermore, a dynamical transition is canonically associated with a redistribution of barrier heights in the energy landscape, which is not determined in this work.  These effects are expected, at most, to modestly perturb the landscape minima.}

\subsection{Spin Representations of the Energy Landscape}

\begin{figure}[t]
\includegraphics[width=\columnwidth]{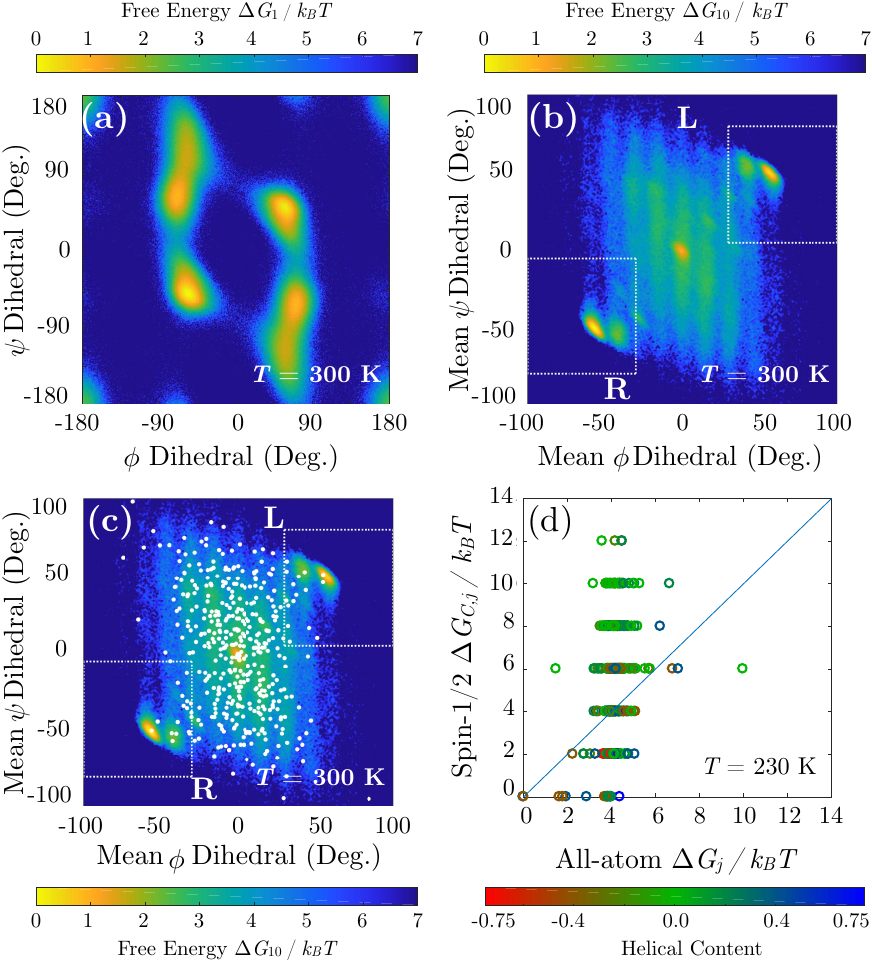} 
\caption{Quantification of the Aib$_{10}$ energy landscape:  (a) free energy landscape $\Delta G_1(\phi, \psi)$ assumed by Aib when incorporated into Aib$_{10}$, derived from the REMD distribution of $(\phi_j,\psi_j)$ backbone dihedral angles; (b) free energy surface $\Delta G_{10}(\bar{\phi}, \bar{\psi})$ for individual Aib$_{10}$ conformers calculated from a distribution of the average backbone dihedrals $(\bar{\phi}_j,\bar{\psi}_j)$; (c) overlay of $k$--means centroids (white dots) with the $\Delta G_{10}(\bar{\phi}, \bar{\psi})$ landscape; and (d) cross--correlation between the $j$--th cluster energy, $\Delta G_j$ determined from REMD simulations, and the $j$--th centroid energy $\Delta G_{C,j}$ calculated using the spin--$1/2$ Hamiltonian $\Delta G_C = \Delta E_\I$ with unit coupling  $(\hat{J}(R,R) = -1.0\,k_B T; \hat{J}(L,R) = 1.0)$. The root mean square error (RMSE) between $\Delta G_j$ and $\Delta G_{C,j}$ is 2.8 $k_B T$, however this is biased toward unstructured states; the RMSE for states lying along the helix--coil transition (Fig. \ref{fig:fitComponent}) is 1.6 $k_B T$.   Energy landscapes were determined using $10^6$ Aib$_{10}$ conformers, extracted from REMD simulations at $T = 230$~K. Boxed areas in (b,c) denote the preferred subregions of dihedral space for left-- and right--handed helices. } \label{fig:twostate}
\end{figure}

\par  The  energy landscape of Aib$_{10}$ -- parameterized by the backbone dihedrals -- reflects the secondary structure of conformational substates and bounds the complexity of coarse--grained  models.  The free energy surface corresponding to a single Aib residue, $\Delta G_1(\phi, \psi)$, is foundational to this classification.  When calculated from the REMD ensemble at $T = 300$~K, this surface contains four primary minima, including left--handed ($\phi \approx -50^\circ , \psi \approx -55^\circ$) and right--handed ($\phi \approx 50^\circ, \psi \approx 55^\circ $) helical regions alongside a pair of broad--shouldered basins (right--handed: $\phi \approx -65^\circ$, $\psi \approx 55^\circ$; left--handed: $\phi \approx 65^\circ$, $\psi \approx -55^\circ$) that define extended conformations (Fig. \ref{fig:twostate}a, \todo{Fig. S9}). REMD simulations reveal a distribution of minima and interconversion barriers (ranging between $4 k_BT$ to $5 k_BT$) that resemble earlier simulations of Aib dynamics, suggesting that our computational approach captures the general condensed phase behavior of Aib.\cite{Grubisic2013, Buchenberg2015}

\begin{figure*}[t]
\bgroup
\setlength\tabcolsep{8.0pt}
\begin{tabular}{ccc}
\includegraphics[]{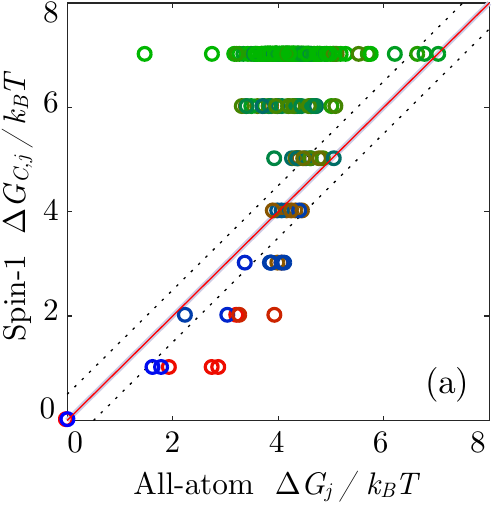} &
\includegraphics[]{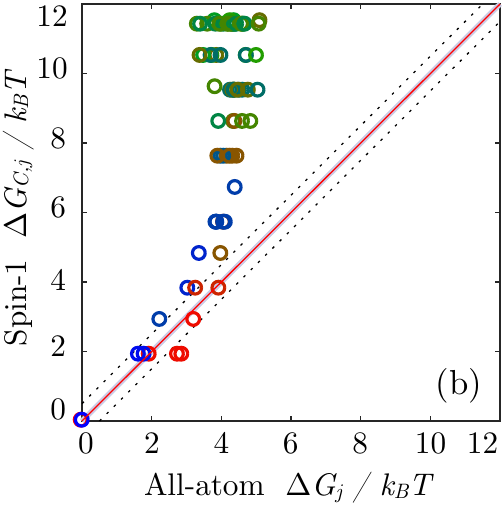} & 
\includegraphics[]{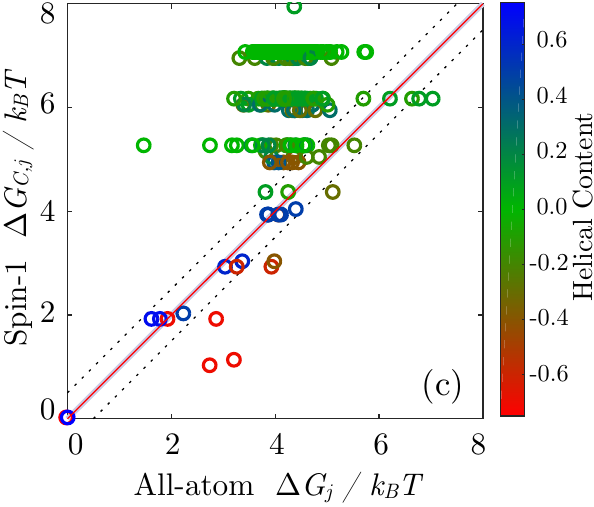}
\end{tabular}
\egroup
\caption{Cross--correlation between the $j$--th cluster energy, $\Delta G_j$ determined from REMD simulations at $T = 230$~K, and the $j$--th centroid energy $\Delta G_{C,j}$ calculated using spin--1 Hamiltonians.  Configurations correspond to (a) $\Delta G_C = \Delta E_\mathcal{I}$, (b) $\Delta G_C = \Delta E_\mathcal{I} + \Delta E_\OO $ and (c) $\Delta G_C = \Delta E_\mathcal{I} + \Delta E_\OO - T \Delta S$.  Points are colored according to their helical content, spanning from right--handed helices (blue) to random coil (green) to left--handed helices (red) (parameters are determined by setting $\hat{J}(L,L) = \hat{J}(R,R) = -1.0\,k_B T$ and $\hat{J}(L,R) = 1.0\,k_B T$ (See SM, Section 1), and employing a systematic fitting procedure to find $K_U = 0.9\,k_B T$, $s = 0.9$).  Dashed lines denote a range within $\pm 0.5\, k_BT$ of the diagonal.  Fitting was performed by maximizing the number of clusters ($N_\text{fit} = 26$) for which  $\vert \Delta G_{C,j} - \Delta G_j\vert \leq 0.5 \, k_B T$ and simultaneously minimizing $\vert \Delta G_{C,j} - \Delta G_j\vert$, affording an RMSE of 0.3~$k_BT$ for fitting points (overall RMSE = 2.4~$k_B T$).  This corresponds to 15 \% of the spin--1/2 RMSE for clusters lying along the helix--coil transition (Fig. \ref{fig:twostate}d). We note that the solvent and entropic terms are both physically tied to the three-state model, as they reflect behavior of the unstructured state. Therefore, they are not included in the spin--1/2 model of Fig.~\ref{fig:twostate}d.} 
\label{fig:fitComponent}
\end{figure*} 
\par In a similar manner, it is straightforward to derive a free energy surface $\Delta G_{10}(\bar{\phi}, \bar{\psi})$ for entire Aib$_{10}$ peptides in terms of the mean backbone dihedrals  $\bar{\phi}_j = (N-2)^{-1} \sum_{\alpha=2}^{N-1} \phi_{j,\alpha}$ and $\bar{\psi}_j = (N-2)^{-1} \sum_{\alpha=2}^{N-1} \psi_{j,\alpha}$, where the index $\alpha$ runs over residues within the $j$--th Aib$_{10}$ configuration.  Summation is restricted to interior residues as the highly--flexible terminal sites can only be assigned a single dihedral parameter.  The resulting landscape is more complex than that of a single Aib residue, containing helical substates that are connected by a near--continuum of weakly structured intermediate configurations (Fig.~\ref{fig:twostate}b).  This architecture is well mapped by $k$--means clusters, with centroids that are both localized in minima of $\Delta G_{10}(\bar{\phi}, \bar{\psi})$ and diffusely distributed throughout the interstitial parameter space (Fig. \ref{fig:twostate}c). 

\par\todo{While this energy profile is consistent with earlier simulations, the preceding efforts identified fewer states within in the energy landscape.\cite{Buchenberg2015}  This deviation is likely associated with the PCA--based dimensional reduction employed by other authors.  In this case, PCA component vectors were shown to convolve the $\phi$ and $\psi$ backbone dihedrals with undetermined lower--weight parameters.  The resulting landscape corresponds (approximately) to a configuration in which conformations from our $k$--means clusters are averaged according to their backbone dihedrals within overlapping neighborhoods (c.f. Fig. \ref{fig:twostate}).  This redistribution and averaging affords a smoothed map of conformational space, while impeding detection of nuanced details that are captured by our clustering.  In a similar manner, the landscape given by $\Delta G_{10}$ contains numerous minima that are better differentiated by the geometric $(\phi,\psi)$ backbone dihedrals than the admixed parameters resulting from PCA.  The marginalization of fine landscape features with certain order parameters is a well--known complication of dimensional reduction \cite{Krivov2004, Krivov2006, Best2010, Wales2015} though any given pair of these parameters may be related through a well--defined scaling transformation.\cite{Best2010} }

\par The Aib$_{10}$ landscape contains numerous conformations (70.8 \% of the ensemble) that lie outside the basins dominated by left-- and right--handed helical character (Fig.~\ref{fig:twostate}; \todo{Fig. S9}). At first glance, this would appear to preclude a two state model, even when restricted to core regions of the Aib helix.  To test this assumption, the centroids at $T = 230$~K were given a binary classification by setting $\sigma_\alpha = R$ when $\psi_\alpha \leq -\phi_\alpha$ and $\sigma_\alpha = L$ when $\psi_\alpha \geq -\phi_\alpha$, following an earlier proposal.\cite{Buchenberg2015}  Using this ensemble, the centroid energies calculated using the spin--$1/2$ Hamiltonian $\Delta G_{C,j} = \Delta E_\I$ exhibit extremely weak correlation (particularly for structured conformers ) with the energies $\Delta G_j$ derived from the REMD landscape, indicating that unstructured configurations are critical to constructing a simplified model of Aib$_{10}$ dynamics. \todo{ This observation is underscored by theoretical investigations of other bistable helical foldamers, which note the importance of unstructured states to either primary or secondary pathways for helix chirality inversion.\cite{Moradi2009,Chakrabarti2011,Chakraborty2017}  }

\par  A more meaningful classification scheme becomes apparent when introducing an unfolded coil ($U$) configuration, leading to a three-state, spin--1 representation of the energy landscape.   To implement this approach, every cluster centroid may be encoded as a spin configuration between residues $i$ and $i+1$ using the function $\text{hcx}[\{\mathbf{x}_i\}, \{\mathbf{x}_{i+1}\}]$ (Eq. \ref{hcx}).  These data may then be employed to compute the the energy $\Delta G_{C,j}$ of the $j$--th centroid, and compared directly to with the corresponding cluster energy $\Delta G_j$ from the all--atom replica ensemble.

\par  The simplest three-state model for Aib$_{10}$ employs only the spin--spin Hamiltonian term $\Delta G_C = \Delta E_\I$ to quantify interactions between structural motifs.  This model demonstrates modest agreement with the REMD cluster distribution at low energies, however, a pronounced deviation is observed for high--entropy states in which random coil character is dominant (Fig. \ref{fig:fitComponent}a).  The inclusion of a correction $\Delta G_C = \Delta E_\I + \Delta E_\OO$ that disfavors solvent--exposed coils strengthens this correspondence for a number of clusters in the well--folded region ($\Delta G_j, \Delta G_{C,j} \leq 4 k_B T$), particularly for states lying within the right--handed funnel (Fig. \ref{fig:fitComponent}b).  While this term accommodates the short helical segments that occur within the bulk of a helix, or the fraying terminal regions associated with the canonical helix--coil transition, there is a dramatic overestimation of solvation penalties for centroids with low helical content.  The introduction of an entropic term $\Delta G_C = \Delta E_\I + \Delta E_\OO - T \Delta S_\R$ strengthens this correspondence (Fig. \ref{fig:fitComponent}c), and may be rationalized as a form of energy--entropy compensation that corrects for the loss of inter--helical hydrogen bonds and helix dipole reinforcement.   Nonetheless, notable deviations persist for a series of states lying within the low--energy regime, as well as within the large high entropy cluster consisting largely of unstructured configurations.  REMD simulations indicate that a $p\Delta V$ term would shift the highest lying centroid energies by approximately 0.7 $k_B T$ (see the SM, Fig. S12). However, the on-site and entropic terms already capture part of this correction due to the fitting. We thus neglect this correction.

\begin{figure}[b]
\includegraphics[width=\columnwidth]{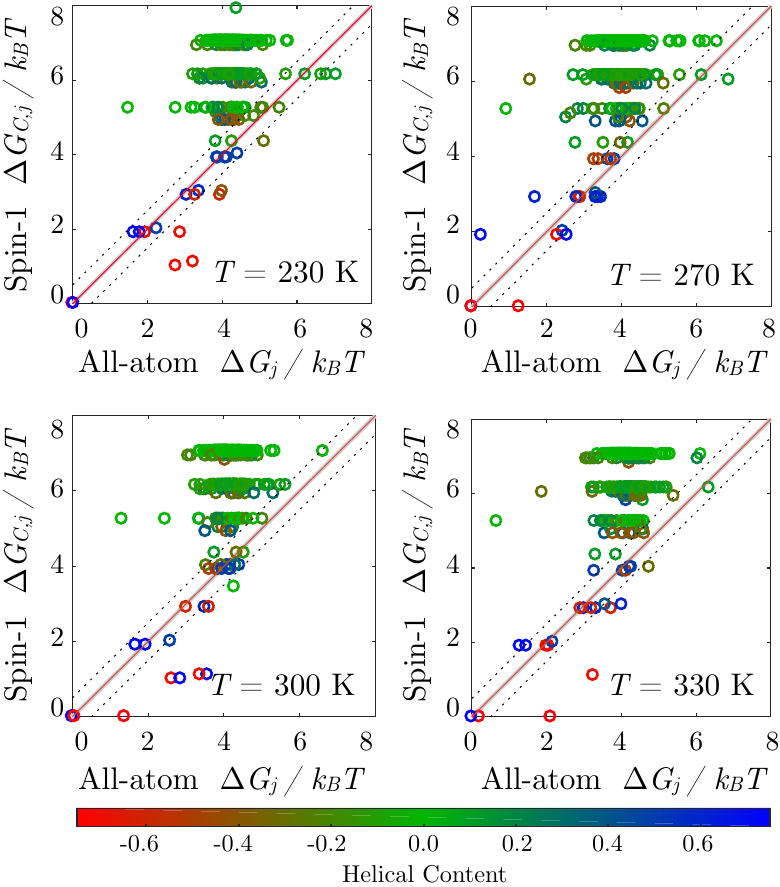}
\caption{Temperature dependence of cross--correlation between the $j$--th cluster energy $\Delta G_j $ and the $j$--th centroid energy $\Delta G_{C,j}$ calculated using the full spin--1 model $\Delta G_C = \Delta E_\I + \Delta E_\OO - T\Delta S$.  Parameters are defined at each temperature so that $\hat{J}(R,R) = 1.0~k_BT$, $\hat{K}(U) = 0.9~k_BT$, and $s = 0.9$, following Fig. \ref{fig:fitComponent}. Dashed lines denote the fitting range within $\pm 0.5\, k_BT$ of the diagonal (overall RMSE$_{230 \text{K}}$ = 2.4 $k_B T$; fit RMSE$_{230 \text{K}} = 0.3~k_BT$ (26 states); overall RMSE$_{270 \text{K}}$ = 2.6 $k_B T$; fit RMSE$_{270 \text{K}} = 0.3~k_BT$ (19 states);  overall RMSE$_{300 \text{K}}$ = 2.5 $k_B T$; fit RMSE$_{300 \text{K}} = 0.3~k_BT$ (25 states); overall RMSE$_{330 \text{K}}$ = 2.4 $k_B T$; fit RMSE$_{230 \text{K}} = 0.3~k_BT$ (19 states)).} 
\label{fig:fitTemp}
\end{figure}

\par It is natural to ask if these deviations are unique to a particular simulation temperature.  To assess this possibility, an identical set of calculations was performed by transferring these parameters, and the full spin model (Eq. \ref{fullmodel}), to four different environmental conditions (Fig. \ref{fig:fitTemp}).  A similar pattern of deviations between $\Delta G_{C,j}$ and $\Delta G_j$ is observed at each temperature, with a comparable quality of fit, suggesting a surprising degree of transferability for this model.  The consistency of this behavior is highly suggestive of a systematic deviation between the spin model and all--atom simulations.

\par
\begin{figure}[t]
\includegraphics[width=\columnwidth]{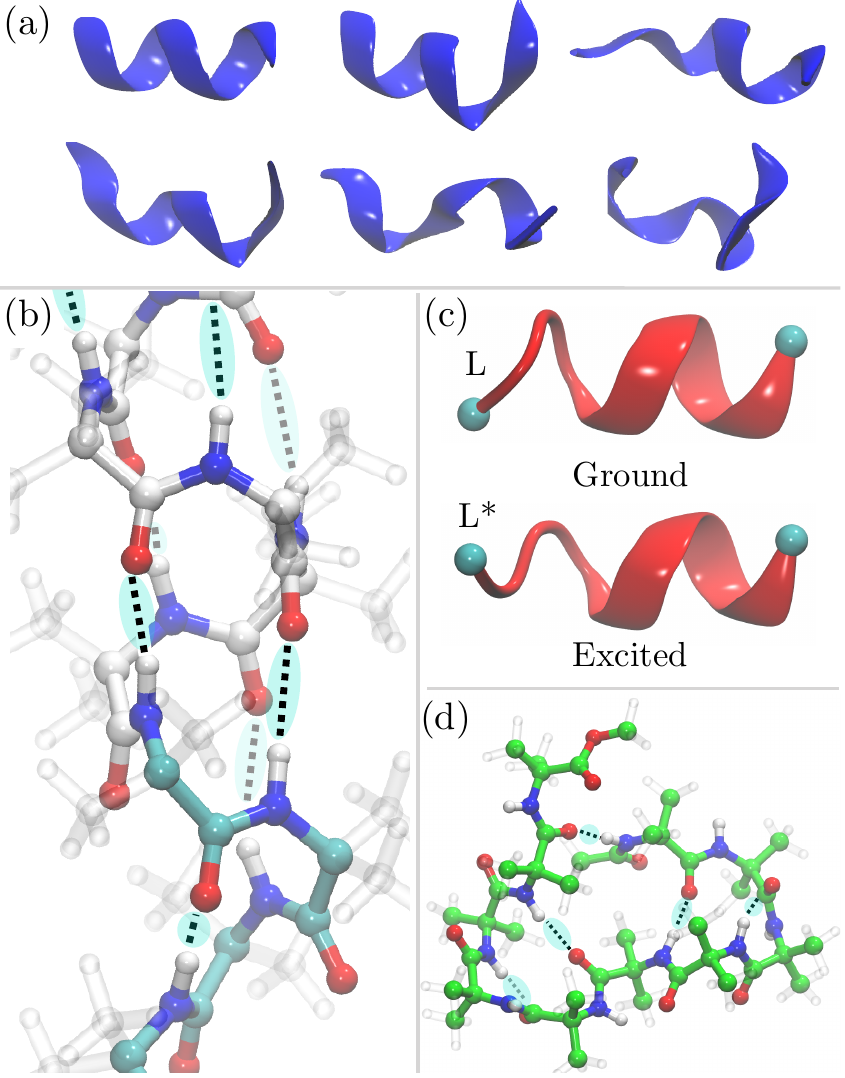}
\caption{Structural elements of the Aib$_{10}$ energy landscape: (a) selected centroids exhibiting tight cross--correlation between the full spin model (Eq. \ref{fullmodel}) and REMD simulations at $T = 230$~K, demonstrating a robust helix--coil transition; (b) cluster demonstrating both $\alpha$-- and $3_{10}$--helical domains, shown in white (on top) and cyan (on bottom), respectively.  Hydrogen bonding patterns are indicated by dotted lines and diffuse shading; (c) comparison between `ground' and `excited' state conformers sharing the same spin assignment; (d) anomalous cluster demonstrating an incommensurate hydrogen bonding pattern.} 
\label{fig:structures}
\end{figure}

 To dissect the origin of this behavior, it is helpful to examine a series of centroids that exhibit tight agreement with replica exchange clusters.  An ideal set is afforded by the right--handed helical funnel within the $T = 230$~K ensemble (Fig. \ref{fig:structures}a).  Cursory analysis of these states reveals a classical helix--coil transition, proceeding from a well--formed $\alpha$--helical native state to a series of higher energy conformers characterized by fraying of the helix termini and the ultimate unwinding of the segment.  The robust fit for this series is consistent with the enhanced stability of $\alpha$--helical populations within Aib foldamers at low temperatures.  Furthermore, the majority of these centroids retain a degree of helicity, suggesting that all but the highest--energy configurations in this series correspond to folds in which a helix has already nucleated.

\par It is notable that low energy clusters --- exhibiting a reasonable correspondence between the spin--1 model and replica exchange simulations --- are dominated by structures with substantial $\alpha$--helical character (Fig. \ref{fig:helixContent}).  Higher energy clusters are dominated by a $3_{10}$--fold when nontrivial helicity is present.  Manual inspection of poorly fit clusters reveals that many contain at least a small $3_{10}$--helical twist, either in isolation (high--entropy cluster) or abutting an $\alpha$--helical segment through a small unstructured linker (states below the diagonal).  Owing to their distinct physical characteristics, these $3_{10}$--helical folds constitute a distinct helical population that coexists alongside the dominant $\alpha$--helical distribution (Fig. \ref{fig:structures}b).  A more physically accurate model might accommodate two distinct helical populations, forming a five--state representation.  The prevalence of $3_{10}$--helical configurations in short stretches underscores a role as nucleation sites for $\alpha$--helices, consistent with their purported role in macromolecular structure formation and with traditional analytical models for the helix--coil transition.\cite{Zimm1959,Lifson1961}

\par An additional family of deviations is associated with clusters that classify into the same spin encoding, yet differ conformationally within the REMD ensemble.  These structures are generally related through a localized conformational distortion, which preserves the overall secondary structure yet places one conformer into a higher--energy configuration (Fig. \ref{fig:structures}c).  These energetically `excited' states were anticipated in earlier PCA analyses\cite{Buchenberg2015} and fall outside the scope of simple spin--based models (one needs further fine graining, i.e., a higher dimensional spin, to account for them), attesting to the importance of all--atom simulations for characterizing macromolecular energy landscapes.  Underscoring this point, several of these configurations are highly populated ($\Delta G_j$ ranging between  2 $k_B T$ and 3\, $k_B T$) and are associated with alterations in the terminal domains of the helix.

\par One final anomaly deserves further discussion.  The atypical cluster located at $\Delta G_j \approx 1.5\, k_B T$ and $\Delta G_{C,j} \approx 6\, k_B T$ is accompanied by several similarly folded counterparts in both left and right helical funnels, and deviates strongly from a helical fold while retaining a well--defined secondary structure (Fig. \ref{fig:structures}d).   These folds contain hydrogen bonding patterns consistent with single turns of $3_{10}$-- and $\alpha$--helical character.  Nonetheless, consecutive backbone hydrogen bonds from the $i$--th residue to  residue $i+3$ ($3_{10}$--helix) and from the $i$--th residue to residue $i+4$  ($\alpha$--helix) lie out of registry and thus a long--distance helical structure fails to form.  This behavior may be unique to the Aib$_{10}$ model, which exceeds the length of prior experimental constructs  and lacks the bulky flanking groups present in synthetic helices.  It is unlikely that this unconventional fold plays a role in the experimentally observed \LRC\, interconversion due to constraints from the helix termini or the surrounding membrane environment.  Nonetheless, simulations of the helix--coil transition in polyalanine indicate the presence of alternate folds that reside within shoulders of the folding funnel.\cite{Levy2001}  It is unclear if these clusters represent similar behavior, or if they correspond to a unique transition pathway between left--handed and right--handed funnels.  While the latter possibility is less probable, it may only be excluded by mapping a kinetic network for these energy landscapes.

\section{Conclusions}

\par Taken together, the observations herein underscore the complexity of seemingly simple macromolecular energy landscapes. From an analytical perspective, a three--state, spin--1 model can accommodate the general structural motifs present within the conformational ensemble of Aib$_{10}$ --- corresponding to left--handed helices, right--handed helices, and unstructured coils.  Nonetheless, the presence of competing $\alpha$-- and $3_{10}$--helical subpopulations limits the scope of this approach.  More complex five--state models can be constructed following Ising or Potts Hamiltonians, however, the presence of high--energy (`excited') and low--energy (`ground') conformational states with identical spin encodings suggests that increasing the dimensions of the model will be met with diminishing returns. \todo{If a high resolution picture is required for the entire state spectrum, Markov state models\cite{Noe2008} or transition path representations\cite{Wales2002, Wales2004,Adjanor2006, Picciani2011,Cameron2014} may afford a more efficacious --- yet costly --- approach.  }

\par Despite these limitations, the construction of simplified (and, in our case, analytical) models helps to reveal key features of the energy landscape.  These considerations extend to techniques for reduction of dimensionality --- while PCA analysis reveals major minima, an approach based around replica exchange and unsupervised clustering captures states that might otherwise be masked when the resolution of simulation data is low.  The importance of excited states and competing folds, as revealed through the spin--1 model, attests to the importance of careful landscape quantification.  A judicious choice of methods is essential, balanced by a tradeoff between computational cost and resolution required of the resulting coarse--grained representation.  \\

On a more sobering note, it is widely known that force-field dependence of the free energy landscape -- and secondary structure in particular -- is a constant source of figurative, and sometimes literal, frustration\cite{Freddolino2009}. For Aib in particular, we find that $\alpha$--helical content dominates at low energy (and $3_{10}$ at higher energy). The funnel crossover energetics are in agreement with other force fields, but the ratio of $\alpha$ to $3_{10}$ content for those cases is not available. Furthermore, experimental data and other MD simulations are not directly transferrable to the longer, unmodified, Aib polypeptide that we examine. \todo{Nevertheless, the inclusion of unstructured regions vastly improves coarse graining, irrespective of the force field -- in this manner we consider Aib$_{10}$ to be a model system for other helical macromolecules.\cite{Moradi2009,Chakrabarti2011,Chakraborty2017}  }

 This gives a general lesson for coarse-graining, whether into discrete structural states or at the atomic scale: Full characterization of the energy landscape for protein fragments and polypeptides is possible. A comparison of  structural features between all-atom, atomically coarse-grained, and discrete representations can therefore give a strong and quantitative assessment of what is physically occurring in these models -- in our case, a 0.3 $k_B T$ deviation of coarse grained states between 0~$k_BT$ and 4 $k_B T$ (this is in addition to constraints that ensure a correspondence with thermodynamic parameters and other data).  In doing so, one can to pinpoint deviations, determine symmetries and asymmetries and, all-in-all,  see what these complex atomic models are really yielding.

\section{Supplementary Material}

\par The supplementary material contains a general evaluation of spin--1/2 and spin--1 model parameters.

\section{Acknowledgements}

J. E. E. acknowledges support under the Cooperative Research Agreement between the University of Maryland and the National Institute for Standards and Technology Center for Nanoscale Science and Technology, Award 70NANB14H209, through the University of Maryland.  K. A. V. was supported by the U.S. Department of Energy through the LANL/LDRD Program. \todo{ Computing resources were made available through the Los Alamos National Laboratory Institutional Computing Program, which is supported by the U.S. DOE National Nuclear Security Administration under contract no. DE-AC52-06NA25396 as well as the Maryland Advanced Research Computing Center (MARCC).}

\vspace{\baselineskip}

\section{Appendices}

\subsection{Definition of Helical Order Parameters}

\par Helical content is quantified \cite{Iannuzzi2003, Alemani2010, Fiorin2013} using a function $\text{Hlx}[\{\mathbf{x}_i\}]$ that scores a given peptide configuration $\{\mathbf{x}_i\}$ based on (i) conformity to the angle formed by consecutive $\alpha$--carbons in an ideal helix and (ii) consistency with an ideal hydrogen bonding arrangement for either an $\alpha$--helix or a $3_{10}$--helix:  

\begin{multline}
\text{Hlx}[\{\mathbf{x}_i\}] = \frac{1}{2(N-2)} \sum_{i=1}^{N-2} \text{Ang}(\mathbf{x}_{\alpha,i}, \mathbf{x}_{\alpha,i+1}, \mathbf{x}_{\alpha,i+2}) \\
+ \frac{1}{2(N-m)} \sum_{i=1}^{N-m} \text{Hb}(\mathbf{x}_{\text{O},i}, \mathbf{x}_{\text{N},i+m})
\end{multline} 

\noindent where $m = 3$ for a $3_{10}$--helix, $m=4$ for an $\alpha$--helix, $\theta_0 = 90^\circ$ reflects the angle formed by three consecutive $\alpha$--carbons and $\Delta \theta_\text{tol} = 15^\circ$ defines an acceptance tolerance for deviations from an ideal helix.  In this case, $\mathbf{x}_{\alpha,i}$, $\mathbf{x}_{\text{O},i}$, and $\mathbf{x}_{\text{N},i}$ denote, respectively, the $\alpha$--carbon, amide oxygen, and amide nitrogen coordinates for the $i$--th residue. The angular deviation function $\text{Ang}(\mathbf{x}_{\alpha,i}, \mathbf{x}_{\alpha,i+1}, \mathbf{x}_{\alpha,i+2})$ is defined as

\begin{widetext}
\begin{equation}
\text{Ang}(\mathbf{x}_{\alpha,i}, \mathbf{x}_{\alpha,i+1}, \mathbf{x}_{\alpha,i+2}) = \frac{1 - [\theta(\mathbf{x}_{\alpha,i}, \mathbf{x}_{\alpha,i+1}, \mathbf{x}_{\alpha,i+2}) - \theta_0]^2 / (\Delta \theta_\text{tol}))^2}{1 - [\theta(\mathbf{x}_{\alpha,i}, \mathbf{x}_{\alpha,i+1}, \mathbf{x}_{\alpha,i+2}) - \theta_0]^4 / (\Delta \theta_\text{tol}))^4}
\end{equation}
\end{widetext}

\noindent where $\theta(\mathbf{x}_{\alpha,i}, \mathbf{x}_{\alpha,i+1}, \mathbf{x}_{\alpha,i+2})$ is the angle formed by consecutive $\alpha$--carbons and the hydrogen bonding contribution is quantified through 

\begin{equation}
\text{Hb}(\mathbf{x}_{\text{O},i}, \mathbf{x}_{\text{N},i+R}) = \frac{1 - [\vert \mathbf{x}_{\text{O},i} - \mathbf{x}_{\text{N},i+R}\vert / d_0]^4}{1 - [\vert \mathbf{x}_{\text{O},i} - \mathbf{x}_{\text{N},i+R}\vert / d_0]^6} .
\end{equation}

\noindent The orientation of a given peptide configuration $\{\mathbf{x}_i\}$  is assigned using the function

\begin{equation}
\text{Hcx}[\{\mathbf{x}_i\}] = \sum_{i=1}^{N-1} \text{hcx}[\{\mathbf{x}_i\}, \{\mathbf{x}_{i+1}\}],
\end{equation}

\noindent where the pairwise helical content is defined in terms of the $(\phi,\psi)$ dihedrals so that

\begin{multline}\label{hcx}
\text{hcx}[\{\mathbf{x}_i\}, \{\mathbf{x}_{i+1}\}] = \\ \begin{cases} 
     1  & (-100^\circ \leq \phi \leq -30^\circ; -80^\circ \leq \psi \leq -5^\circ) \\
      0 & \text{otherwise} \\
      -1 & (100^\circ \geq \phi \geq 30^\circ; 80^\circ \geq \psi \geq 5^\circ) 
   \end{cases}
\end{multline}

\noindent reflects the net right--handed (positive) or left--handed helical content (negative), respectively.

%


\end{document}


\newcommand{\todo}[1] {{\color{red} #1}}

\newcommand{\LRC}{$L \leftrightarrow R$}
\newcommand{\I}{\mathcal{I}}
\newcommand{\E}{\mathcal{E}}
\newcommand{\C}{\mathcal{C}}
\newcommand{\OO}{\mathcal{O}}
\newcommand{\R}{\mathcal{R}}

 \renewcommand{\thefigure}{S\arabic{figure}}

\begin{center}
{\bf Supplementary Information for: \\ ``A Spin--1 Representation for Dual--Funnel Energy Landscapes''}\\
\vspace{\baselineskip}
Justin E. Elenewski,$^{1,2}$ Kirill A. Velizhanin,$^3$ Michael Zwolak$^1$\\
\vspace{0.5\baselineskip}
$^1$\emph{Center for Nanoscale Science and Technology, \\ National Institute of Standards and Technology, Gaithersburg, MD 20899, USA}\\
\vspace{0.5\baselineskip}
$^2$\emph{Maryland Nanocenter, University of Maryland, College Park, MD 20742, USA}\\
\vspace{0.5\baselineskip}
$^3$\emph{Theoretical Division, Los Alamos National Laboratory, Los Alamos, NM 87545 USA}
\end{center}

\vspace{5\baselineskip}
\tableofcontents

\clearpage
\section{Fitting Parameters for Spin-1/2 and Spin-1 Models}

\subsection{Spin--1/2 Model}  The symmetrically--coupled spin--1/2 model $\Delta G_C = E_\I$ is straightforward to parameterize since it has only one free parameter $\hat{J}(R, R) = \hat{J}(L, L) = J$ and $\hat{J}(L, R) = \hat{J}(R, L) = -J$.  We restrict discussion to this case, since an asymmetric $\hat{J}$ fails to reproduce key aspects of the REMD energy landscape.   As a general rule, centroid free energies $\Delta G_{C,j}$ calculated using the spin--1/2 model exhibit poor correlation with cluster free energies  $\Delta G_j$ from REMD simulation.  This is due to the absence of an unstructured state, leading to an artificial subdivision of the $(\phi,\psi)$ dihedral space into left-- and right--handed regions.  This limitation makes it difficult to define an optimization target.

\par  A physically meaningful alternative can be derived using optimal fits for the symmetric spin--1 model.  In this case, we can identify a set of $N_\text{fit} = 26$ centroids which deviate from the corresponding cluster energies by less than $ 0.5 \, k_B T$, corresponding to a conventional helix--coil transition.  Any meaningful model for the \LRC\, interconversion must capture this process.  In our fitting protocol, we first minimize the sum--square deviation between centroid free energies  in the spin--1/2 model $(\Delta G_{C,j})$ and from all--atom simulations $(\Delta G_j)$ using states in the aforementioned set

\begin{equation}
F = \sum_{j =1 }^{N_\text{fit}} \vert \Delta G_{C,j} - \Delta G_j\vert^2 
\end{equation}

\noindent and then rank comparable solutions to maximize the number of centroids for which $\vert \Delta G_{C,j} - \Delta G_j\vert \leq 0.5 \, k_BT$.  The results of this method are summarized in Fig. S1 and Fig. S2.

\begin{figure}[h]
\begin{tabular}{|c||c|c|c|}
\hline
$J /k_B T  $       & RMSE(fit) / $k_B T$  & RMSE(ens) / $k_B T$ & $N_{\pm 0.5}$  \\
\hline
\hline
-0.7 & 1.8 & 1.8 & 97 \\
-0.8 & 1.6 & 2.1 & 57 \\
-0.9 & 1.6 & 2.4 & 58 \\
\hline
{\bf -1.0} & {\bf 1.6} & {\bf 2.8} & {\bf 85} \\
\hline
-1.1 & 1.8 & 3.3 & 82 \\
-1.2 & 2.0 & 3.9 & 36 \\
-1.3 & 2.3 & 4.0 & 15 \\
\hline
\end{tabular}
\caption{Fitting parameters for the spin--1/2 model at $T = 230$ K.  Quality of fit is assessed through a root--mean--square error (RMSE) calculated for all $N_\text{fit}$ clusters within the fitting set  $\text{RMSE(fit)} = [N_\text{fit}^{-1} \sum_{j=1}^{N_\text{fit}} \vert \Delta G_{C,j} - \Delta G_{j} \vert^2 ]^{1/2}$  as well as the number $N_{\pm0.5}$ of clusters falling within a free energy window $\vert \Delta G_{C,j} - \Delta G_j\vert \leq 0.5 \, k_BT$.  The RMSE for the full ensemble  $\text{RMSE(ens)} = [N^{-1} \sum_{j=1}^{N} \vert \Delta G_{C,j} - \Delta G_{j} \vert^2 ]^{1/2}$ is provided for reference. The optimal parameter set is designated in {\bf bold.}}
\end{figure}

\vspace{-\baselineskip}
\begin{figure}[h]
\includegraphics[width=6.5in]{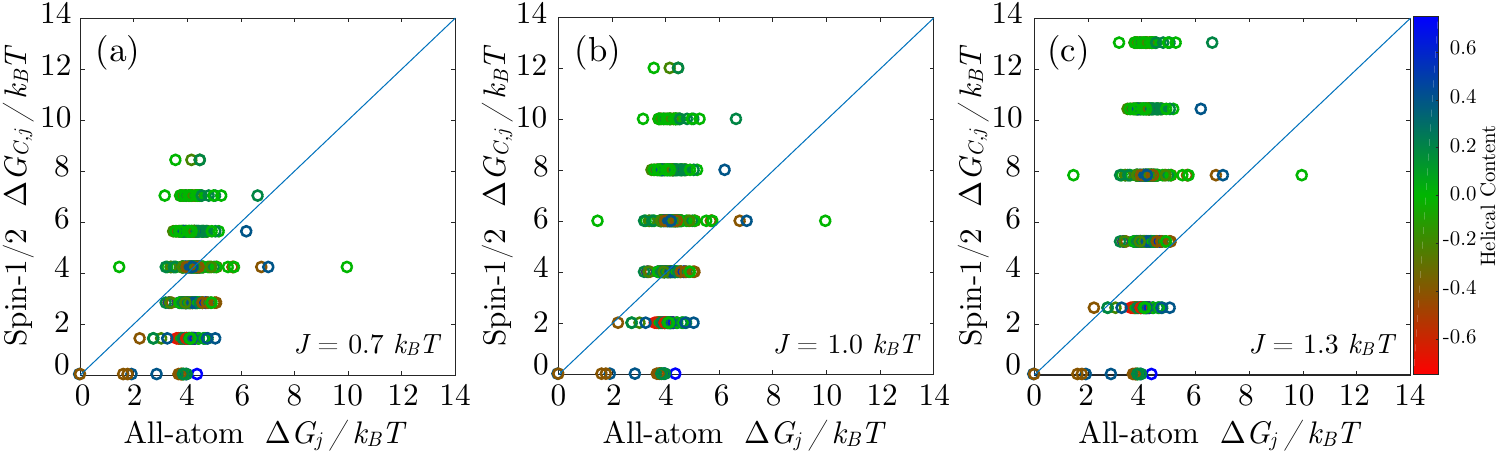}
\caption{Representative fits for the symmetrically--coupled spin--1/2 Hamiltonian model $\Delta G_C = \Delta E_\I$ versus  cluster  energies  $\Delta G_j$ from all--atom REMD simulations at $T = 230$ K.  Datasets correspond to (a) $J(R,R) = 0.7\, k_B T$; (a) an optimal fit of $J(R,R) = 1.0\, k_B T$; and (c) $J(R,R) = 1.3\, k_B T$.  The spin--1/2 model essentially cannot model the REMD data, irrespective of parameters. }
\end{figure}

\clearpage


\subsection{ Spin--1 Model}  

\par The spin--1 model $\Delta G_C = \Delta E_\I + \Delta E_\OO - T\Delta S_\R$ is defined by three independent parameters --- a spin--spin coupling $\hat{J}(R,R) = J$ between adjacent structured ($L/R$) sites, the on--site term for unstructured sites $\hat{K}(U)$ and the site--wise entropy $k_B \cdot s$ assigned to coiled segments.  For simplicity, we will restrict our discussion to a symmetric case $\hat{J}(R,R) = \hat{J}(L,L) = J$; $\hat{J}(R,L) = \hat{J}(L,R) = -J$ where there is no coupling to unstructured segments: $\hat{J}(R,U) = \hat{J}(L,U)$ and $\hat{J}(U,U) = 0$.  This symmetric case, neglecting nucleation terms, is the configuration most physically consistent with all--atom simulations.

  In this case, we can identify a set of $N_\text{fit}$ clusters with $\vert \Delta G_{C,j} - \Delta G_j\vert \leq 0.5 \, k_B T$ that correspond to a conventional helix--coil transition.  Any meaningful model for the \LRC\, interconversion must --- at a minimum --- capture the physics of this process.  In our fitting protocol we first minimize the sum--square deviation between centroid free energies $\Delta G_{C,j}$, calculated using the spin--1 model, and those from all--atom simulations $\Delta G_j$.  This optimization is only performed for states in the aforementioned set.  We simultaneously maximize the number of points in the band $\vert \Delta G_{C,j} - \Delta G_j\vert \leq 0.5 \, k_B T$ to ensure generality of our parameter set.  
  
\par Our optimization is performed as a scan in the spin--spin coupling parameter $J$ and as a systematic search in $(\hat{K}(U), s)$ values.  That is, for each $J \in \{J_1, \dots, J_n\}$ we seek

\begin{equation}
(\hat{K}(U), s) = \text{arg min} \left[ \sum_{j =1 }^{N_\text{fit}} \vert \Delta G_{C,j}(J,\hat{K}(U), s) - \Delta G_j\vert^2 \right]
\end{equation}

\noindent where minimization is performed over a 400 x 400 point grid.  The results of this process are summarized in Fig. S3, and representative fits are presented in Fig. S4.  It is interesting that the optimial parameter set $(J = 1.00\,k_BT, \hat{K}(U)) = 0.9\,k_BT, s = 0.5)$ represents an intermediate case --- for smaller $J$ the fit is dominated by the high entropy cluster of disordered states, while for larger $J$ this is dominated by the low--energy helical conformers ($\Delta G_j \geq 4\, k_BT$).

\begin{figure}[h]
\begin{tabular}{|c||c|c|c|c|c|}
\hline
{\bf  $J / k_B T$ } & $\hat{K}(U) / k_B T$ & $s$ & { RMSE(fit) } / $k_B T$ & {RMSE(ens) } / $k_B T$ & $N_{\pm0.5 }$ \\
 &  & (per site)&  &  & \\
\hline\hline
-0.6 & 0.7 & 0.7 & 1.7 & 0.9  & 217 \\
-0.7 & 0.7 & 0.7 & 1.3 & 0.8 & 151 \\
-0.8 & 0.8 & 0.8 & 1.0 & 1.1 & 84 \\
-0.9 & 0.8 & 0.8 & 0.6 & 1.7 & 39 \\
\hline
{\bf -1.0} & {\bf 0.9} & {\bf 0.9} & {\bf 0.3} & {\bf 2.4} & {\bf 26} \\
\hline
-1.1 & 0.8 & 0.8 & 0.4 & 2.8 & 21 \\
-1.2 & 0.7 & 0.8 & 0.8 & 3.3 & 14 \\
-1.3 & 0.5 & 0.6 & 1.5 & 4.3 & 9 \\
-1.4 & 0.6 & 0.6 & 1.9 & 5.9 & 8 \\
\hline
\end{tabular}
\caption{Fitting parameters for the spin--1 model at $T = 230$ K.  Quality of fit is assessed through a root--mean--square error (RMSE) calculated for all $N_\text{fit}$ clusters within the fitting set  $\text{RMSE(fit)} = [N_\text{fit}^{-1} \sum_{j=1}^{N_\text{fit}} \vert \Delta G_{C,j} - \Delta G_{j} \vert^2 ]^{1/2}$  as well as the number $N_{\pm0.5}$ of clusters falling within a free energy window $\vert \Delta G_{C,j} - \Delta G_j\vert \leq 0.5 \, k_BT$.  The RMSE for the full ensemble  $\text{RMSE(ens)} = [N^{-1} \sum_{j=1}^{N} \vert \Delta G_{C,j} - \Delta G_{j} \vert^2 ]^{1/2}$ is provided for reference. The optimal parameter set is designated in {\bf bold.}}
\end{figure}

\begin{figure}[h]
\includegraphics[width=6.5in]{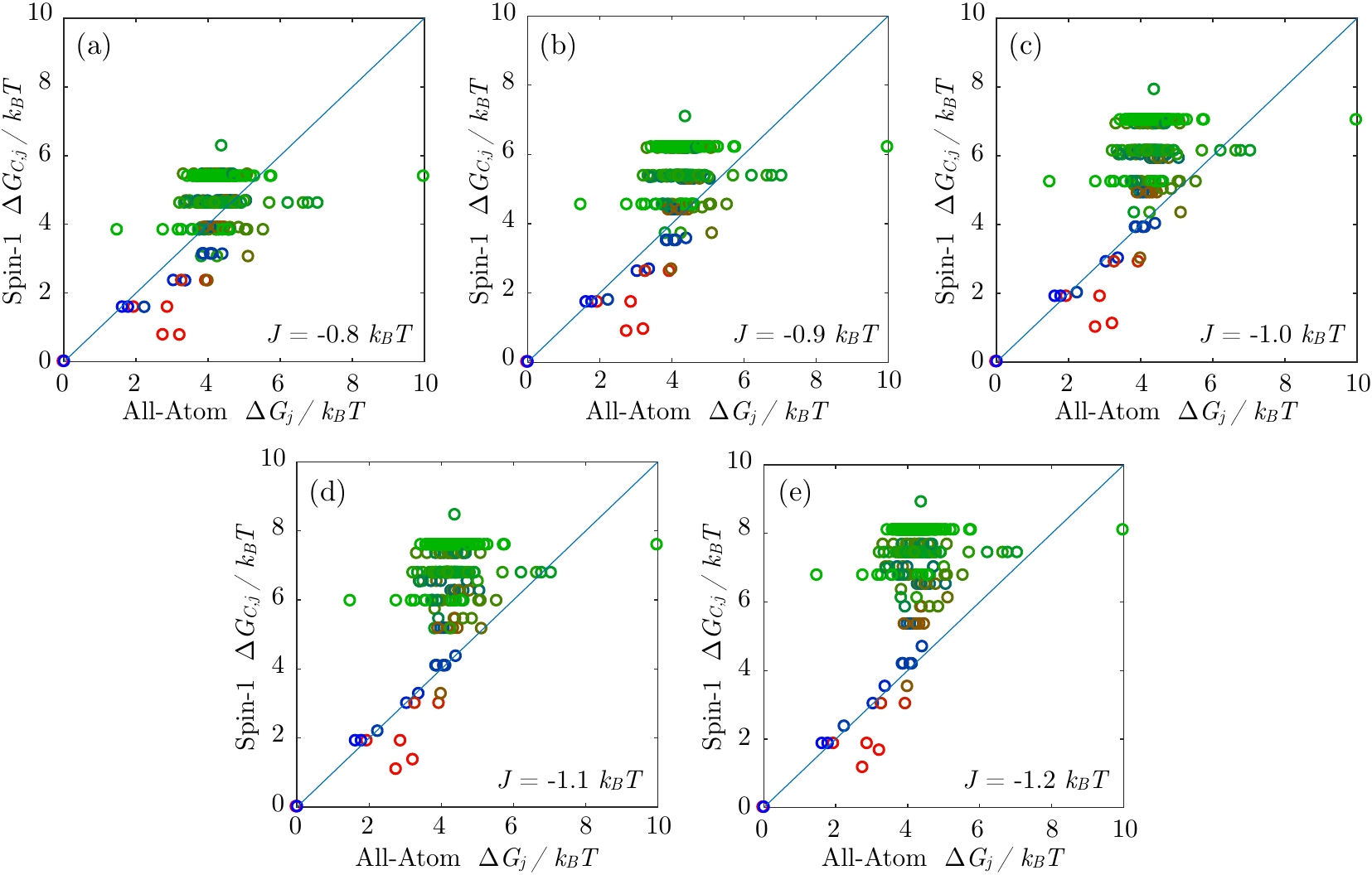}
\caption{Representative fits for the symmetrically--coupled spin--1 Hamiltonian model $\Delta G_C = \Delta E_\I + \Delta E_\OO - T\Delta S_\R$ versus  cluster  energies  $\Delta G_j$ from all--atom REMD simulations at $T = 230$ K.  Datasets correspond to (a) $J(R,R) = -0.8\, k_B T$; (b) $J(R,R) = -0.9\, k_B T$; (c) an optimal fit for $J(R,R) = -1.0\, k_B T$; (d) $J(R,R) = -1.1\, k_B T$; and (e) $J(R,R) = -1.2\, k_B T$.  Parameters in (a,b) are biased toward the disordered, high--energy conformers while (d,e) preferentially fit the low energy region $(\Delta G_j \leq 4\, k_BT$) dominated by helical conformers.  The best fit (c) represents an equally--partitioned physical scenario between these two regions.}
\end{figure}

\clearpage

\section{Asymmetric Two--State (Spin = 1/2) Model: $\hat{J} (R,R) \neq \hat{J}(L,L)$}

\begin{figure}[h]
\includegraphics[width=6.5in]{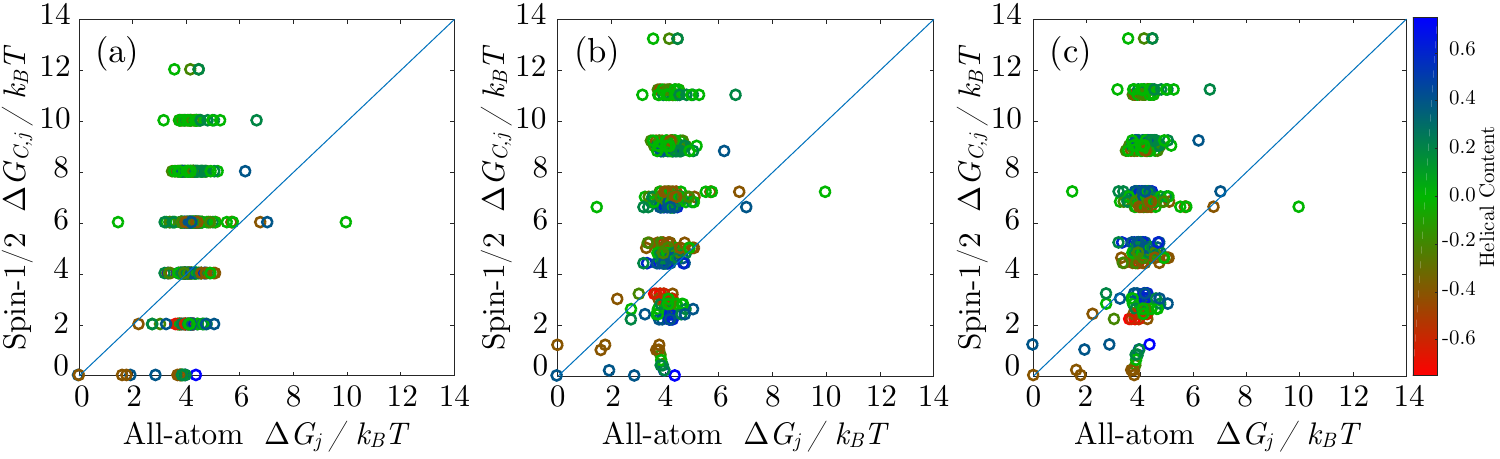}
\caption{The effect of asymmetric spin--spin coupling, $\hat{J}(R,R) \neq \hat{J}(L,L)$, on the spin--1/2 Hamiltonian model $\Delta G_C = \Delta E_\I$, benchmarked against  cluster  energies  $\Delta G_j$ from all--atom REMD simulations at $T = 230$ K.  Calculations are performed under  (a) a reference scenario with symmetric coupling  ($\hat{J}(R,R) = \hat{J}(L,L) = -1.0\,k_BT$); (b) preferential coupling for right--handed conformers ($\hat{J}(R,R) = -1.2\,k_BT; \hat{J}(L,L) = -1.0\,k_BT$); and (c) preferential coupling between left--handed conformers ($\hat{J}(R,R) = -1.0\,k_BT; \hat{J}(L,L) = -1.2\,k_BT$).  The interfacial energy at  domain walls $\hat{J}(R,L) = \hat{J}(L,R) = 1.0\,k_BT$ is identical for all cases.  Note that the centroid energies $\Delta G_{C,j}$ for the lowest energy conformers differ in the presence of an asymmetric coupling -- contrary to all--atom simulations where $\Delta G_j$ is identical for these clusters.  This observation -- that asymmetry breaks the degeneracy (or near degeneracy) in the fully helical states -- indicates that an asymmetric coupling should not be included in our calculations. We do note, however, that the intermediate energy states (the ones between the fully helical states and the disordered states around 4 $k_B T$) suggest that there is, in fact, an asymmetry. This suggest that a next-nearest neighbor model (or longer range) that can simultaneously keep the fully helical state degenerate but give an asymmetry in energy of domain boundaries might further improve the description of the polypeptide with a spin--1 model. Such a range of interaction is physically well-motivated since the helical twist brings amino acids along one turn into close spatially proximity.}
\end{figure}

\clearpage

\section{Asymmetric Three--State (Spin = 1) Model: $\hat{J} (R,R) \neq \hat{J}(L,L)$}

\begin{figure}[h]
\includegraphics[width=6.5in]{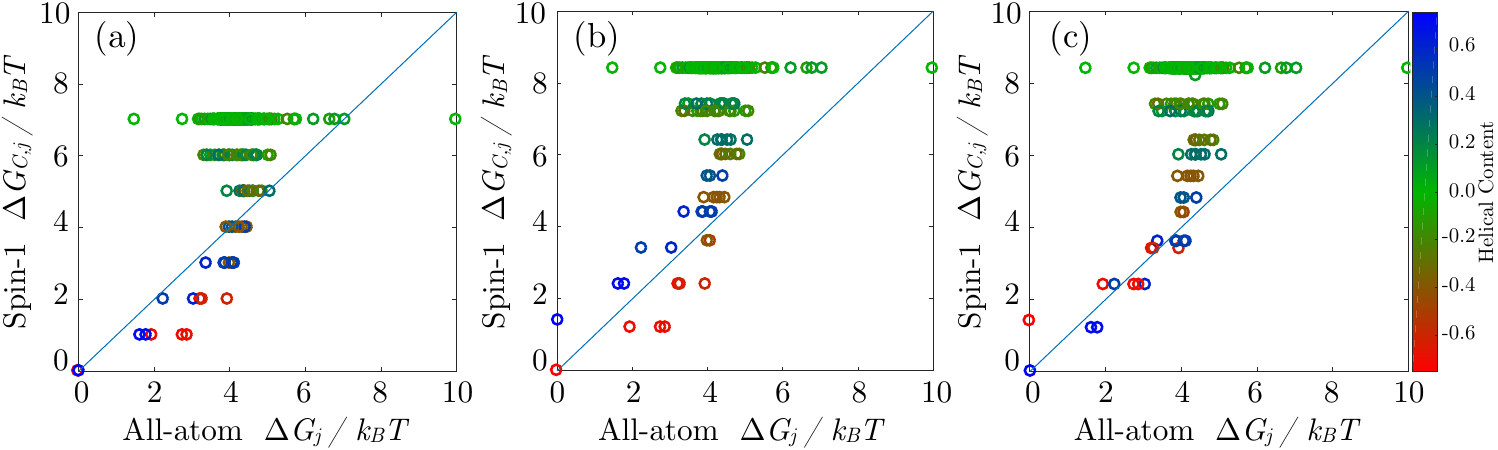}
\caption{ Effect of asymmetric spin--spin coupling, $\hat{J}(R,R) \neq \hat{J}(L,L)$, on the spin--1 Hamiltonian model $\Delta G_C = \Delta E_\I$, benchmarked against  cluster  energies  $\Delta G_j$ from all--atom REMD simulations at $T = 230$ K.  Calculations are performed under  (a) a reference scenario with symmetric coupling  ($\hat{J}(R,R) = \hat{J}(L,L) = -1.0\,k_BT$); (b) preferential coupling for left--handed conformers ($\hat{J}(L,L) = -1.2\,k_BT; \hat{J}(R,R) = -1.0\,k_BT$); and (c) preferential coupling for right--handed conformers ($\hat{J}(L,L) = -1.0\,k_BT; \hat{J}(R,R) = -1.2\,k_BT$).  The interfacial energy at  domain walls $\hat{J}(R,L) = \hat{J}(L,R) = 1.0\,k_BT$ is identical for all cases. Note that the centroid energies $\Delta G_{C,j}$ for the lowest energy conformers differ in the presence of an asymmetric coupling -- contrary to all--atom simulations where $\Delta G_j$ is identical for these clusters.  This observation indicates that an asymmetric coupling should not be included in our calculations.     } 
\end{figure}

\clearpage

\section{Three--State (Spin = 1) Model + Nucleation Correction: $\hat{J} (U,R) = \hat{J}(U,L) \neq 0$}

\begin{figure}[h]
\includegraphics[width=6.5in]{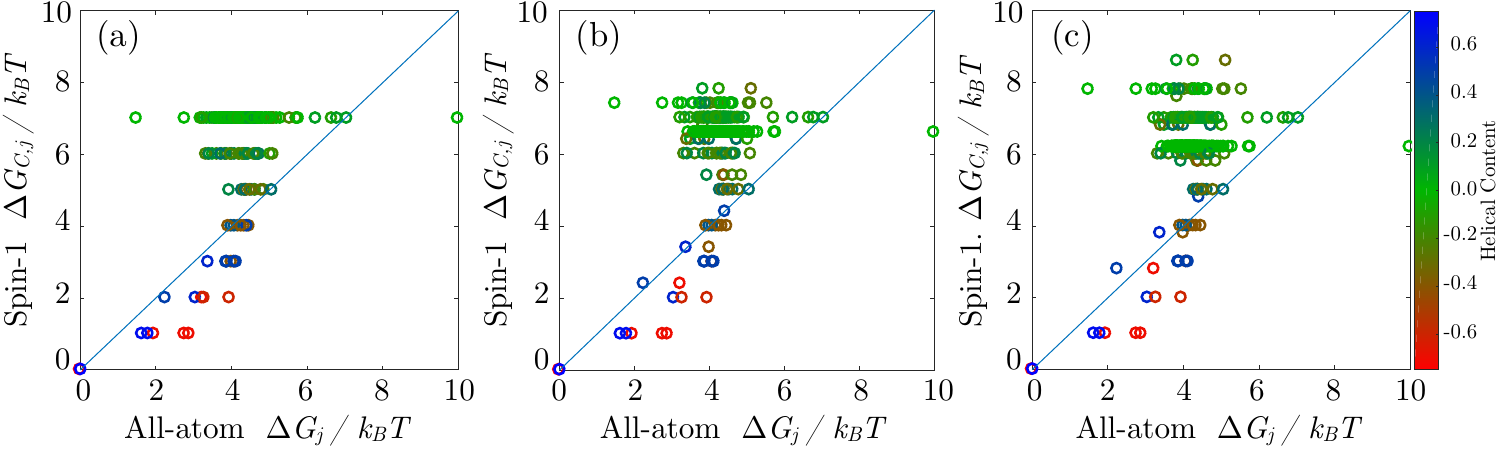}
\caption{ Effect of a nonzero nucleation penalty, $\hat{J}(U,R) = \hat{J}(U,L) \neq 0$, in the spin--1 Hamiltonian model $\Delta G_C = \Delta E_\I$, benchmarked against  cluster  energies  $\Delta G_j$ from all--atom REMD simulations at $T = 230$ K.  Nucleation terms are assumed to be symmetric $\hat{J}(U,R) = \hat{J}(R,U)$.  Calculations are performed under  (a) a reference scenario with symmetric coupling and no nucleation term ($\hat{J}(R,R) = \hat{J}(L,L) = -1.0\,k_BT; \hat{J}(R,L) = 1.0\,k_BT$); (b) reference conditions in (a) plus $\hat{J}(U,R) = \hat{J}(U,L) = 0.2\,k_BT$; (c) reference conditions in (a) plus $\hat{J}(U,R) = \hat{J}(U,L) = 0.4\,k_BT$.  Changes are observed in the distribution of centroid energies $\Delta G_{C,j}$ for states containing coiled segments, shifting a subset of these to higher energies.  This contribution does not markedly improve the correlation between the spin--1 free energies $\Delta G_{C,j}$ and those from all--atom simulations $\Delta G_{j}$.  As a consequence, this nucleation correction is neglected in the main-text calculations.} 
\end{figure}

\clearpage

\section{Helix Terminal Correction: $\hat{J} (\sigma_1,R/L) = \hat{J}(R/L,\sigma_{10}) \neq 0$}

\begin{figure}[h]
\includegraphics[width=6.5in]{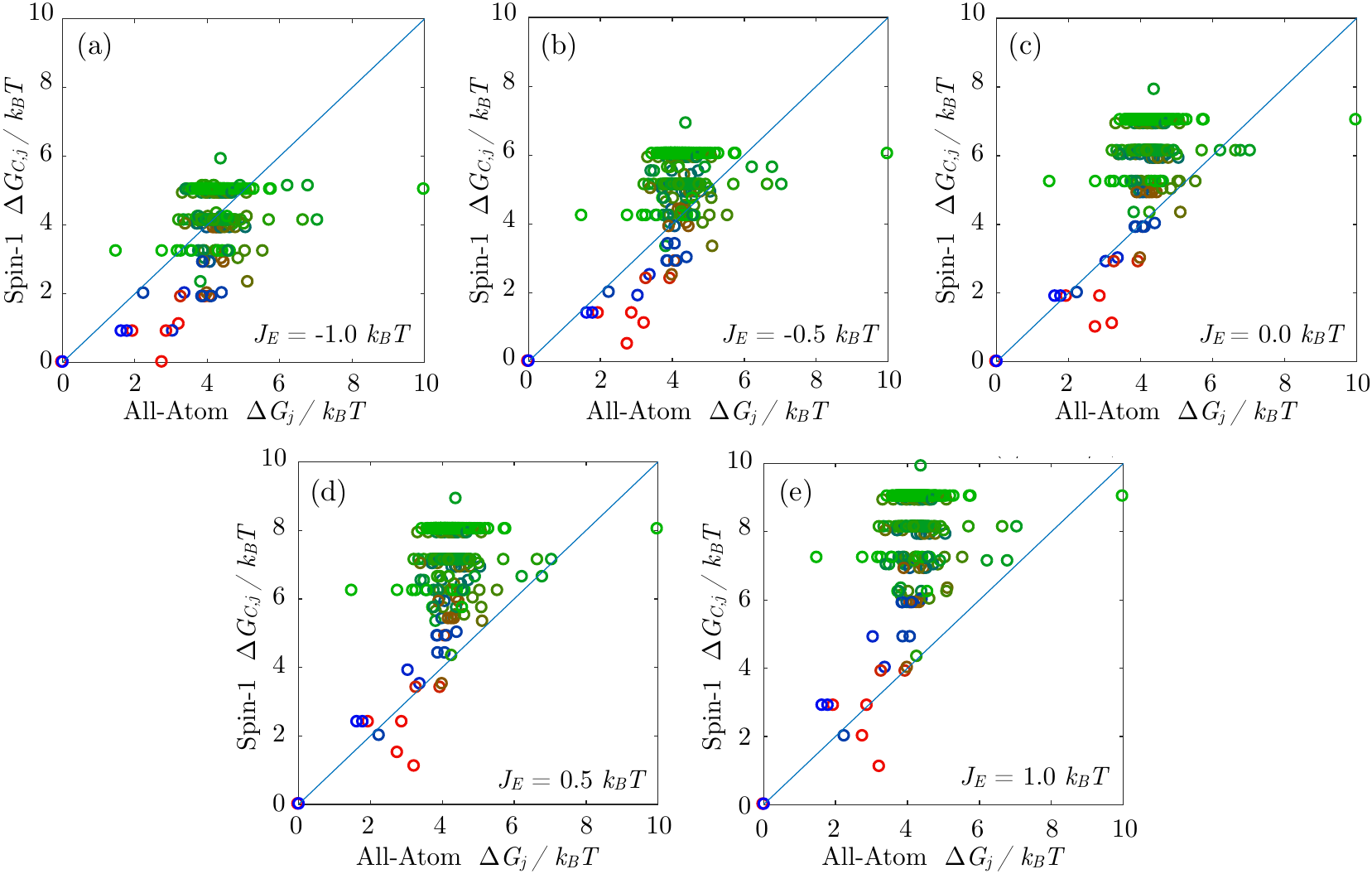}
\caption{ Effect of a terminal correction, introduced as an additional coupling between unstructured helix termini and any structured domains that flank these regions.  The correction $\hat{J}_E(\sigma_1,R/L) = \hat{J}_E(R/L,\sigma_10) = J_E$ is considered for the full spin--1 Hamiltonian model $\Delta G_C = \Delta E_\I +\Delta E_\OO - T\Delta S_\R$ and benchmarked against  cluster  energies  $\Delta G_j$ from all--atom REMD simulations at $T = 230$ K.  Model parameters ($\hat{J}(R,R) = 1.0 \,k_BT; \hat{K}(U) = 0.9\,k_BT; \,k_BT; s = 0.9 $) are adopted from the optimized fit for the spin--1 model in the absence of other corrections.  The terminal correction is evaluated at (a) $J_E = -1.0\,k_BT$; (b) $J_E = -0.5\,k_BT$; (c) as the standard spin--1 model without correction $J_E = 0.0\,k_BT$; (d) $J_E = 0.5\,k_BT$; (e) $J_E = 1.0\,k_BT$, with positive values acting as a nucleation correction and negative values reflecting some local structure to the helix termini.  Due to the abundance of unstructured termini, these corrections largely result in general translations of $\Delta G_C$ values to lower (negative $J_E$) and higher (positive $J_E$) energies.  This correction is  neglected due to the nonspecific nature of its effect.}
\end{figure}

\clearpage

\section{Solvent Accessibility: Helical vs. Compact Conformations}

\begin{figure}[h]
\includegraphics[]{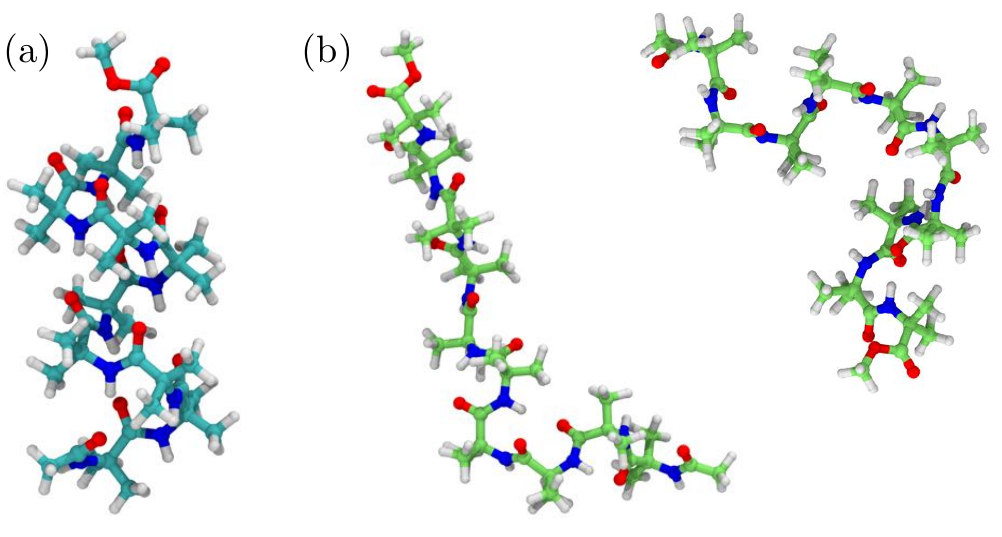}
\caption{Variation in backbone solvent accessibility for Aib$_{10}$ conformers.  Helical conformations (a) restrict solvent access to polar backbone amides, while extended conformers (b) expose these functions to bulk solvent. }
\end{figure}

\clearpage

\section{Three--State (Spin = 1) Model + Asymmetric On--Site Term:  $\hat{K} (R) \neq \hat{K}(L)$}

\begin{figure}[h]
\includegraphics[width=6.5in]{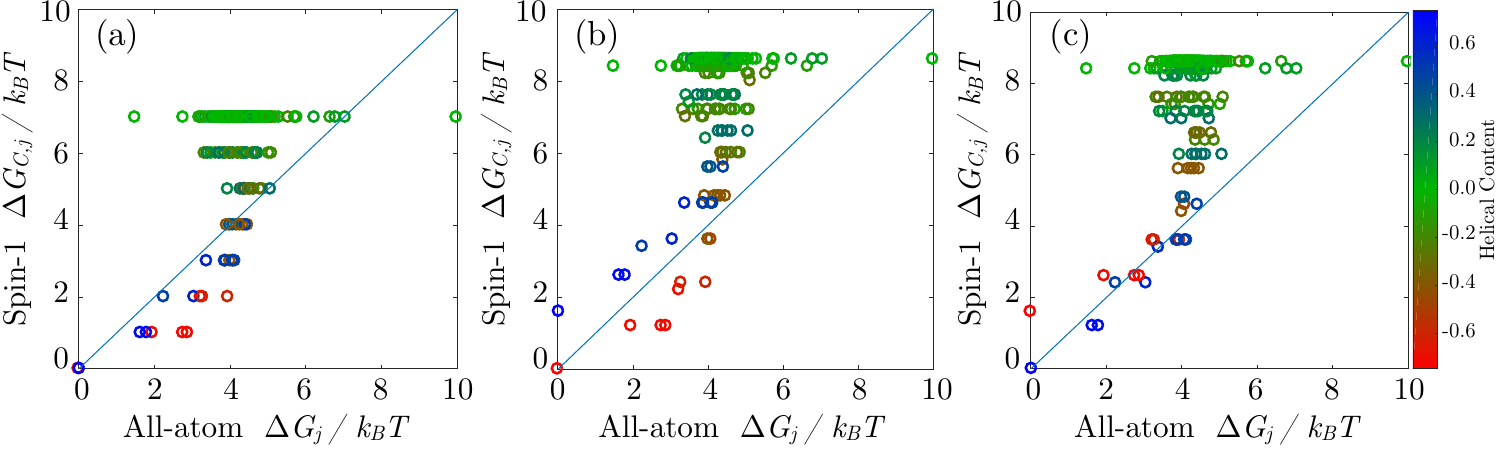}
\caption{ Effect of an asymmetric on--site coupling, $\hat{K}(L/R) \neq 0$, within structured regions of the spin--1 Hamiltonian model $\Delta G_C = \Delta E_\I + \Delta E_\OO$, benchmarked against  cluster  energies  $\Delta G_j$ from all--atom REMD simulations at $T = 230$ K.  Calculations are performed under  (a) a reference scenario with symmetric coupling and no on--site asymmetry ($\hat{J}(R,R) = \hat{J}(L,L) = -1.0\,k_BT; \hat{J}(R,L) = 1.0\,k_BT; \hat{K}(L/U/R) = 0$); (b) reference conditions in (a) plus $\hat{K}(L) = -0.2\,k_BT$, favoring left--handed helices ; (c) reference conditions in (a) plus $\hat{K}(R) = -0.2\,k_BT$, favoring right--handed helices.  Note that the centroid energies $\Delta G_{C,j}$ for the lowest energy conformers differ in the presence of an asymmetric coupling -- contrary to all--atom simulations where $\Delta G_j$ is identical for these clusters.  This observation indicates that an asymmetric on--site coupling should not be included in our calculations.} 
\end{figure}

\clearpage

\section{Entropic Scaling}

\begin{figure}[h]
\includegraphics[width=6.5in]{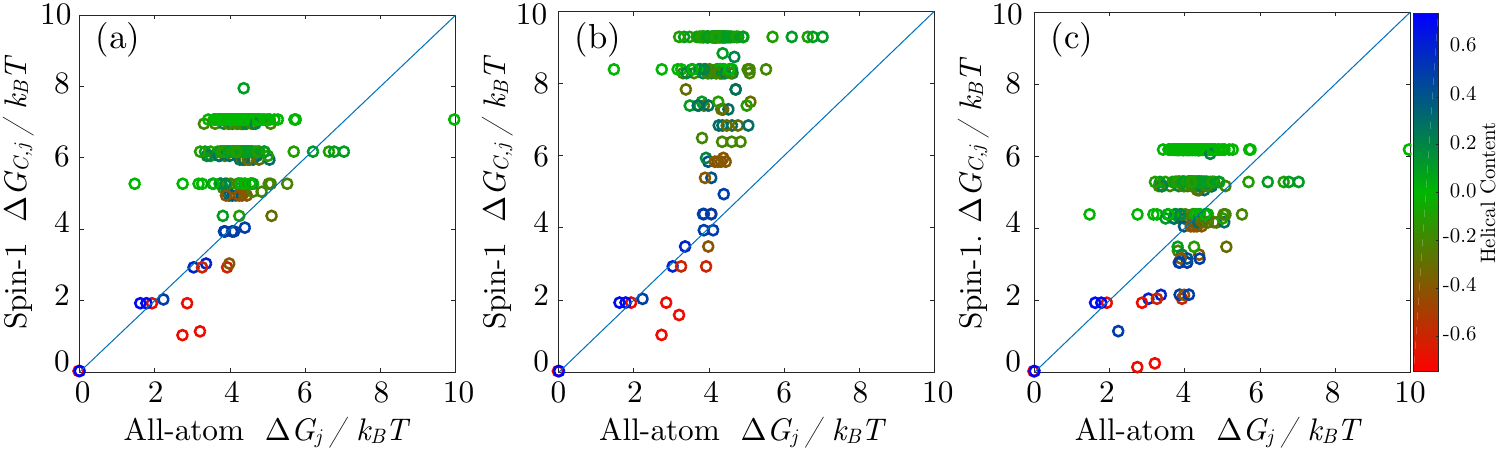}
\caption{ Effect of entropic scaling ($n_i$ versus $n_i -1$) with unstructured chain length $n_i \in \mathcal{C}$ in the spin--1 Hamiltonian model $\Delta G_C = \Delta E_\I + \Delta E_\OO - T\Delta S_\R$.  Models are  benchmarked against  cluster energies  $\Delta G_j$ from all--atom REMD simulations at $T = 230$ K.  Calculations are performed under  (a) the simulation conditions obtained after fitting with $S_\R = k_B \sum_{i \in \mathcal{C}} (n_i - 1) \cdot s$, as described in Fig. 6 of the manuscript ($\hat{J}(R,R) = \hat{J}(L,L) = -1.0\,k_BT; \hat{J}(R,L) = 1.0\,k_BT; \hat{K}(L/U/R) = 0.9\,k_BT$; $s = 0.9$); (b) conditions identical to (a) except with $S_\R = k_B \sum_{i \in \mathcal{C}} n_i \cdot s$ and $s = 0.25$; and (c) conditions identical to (a) except with $S_\R = k_B \sum_{i \in \mathcal{C}} n_i \cdot s$ and $s = 0.5$.  Note that the centroid energy $\Delta G_{C,j}$ of coil--rich clusters decreases too rapidly as a function of $s$ when $n_i$ scaling is employed.  This impedes fitting and further supports the use of an $(n_i -1)$ scaling term in the spin--1 model. } 
\end{figure}

\clearpage

\section{Three--State (Spin = 1) Model: Pressure Correction}

\begin{figure}[h]
\includegraphics[]{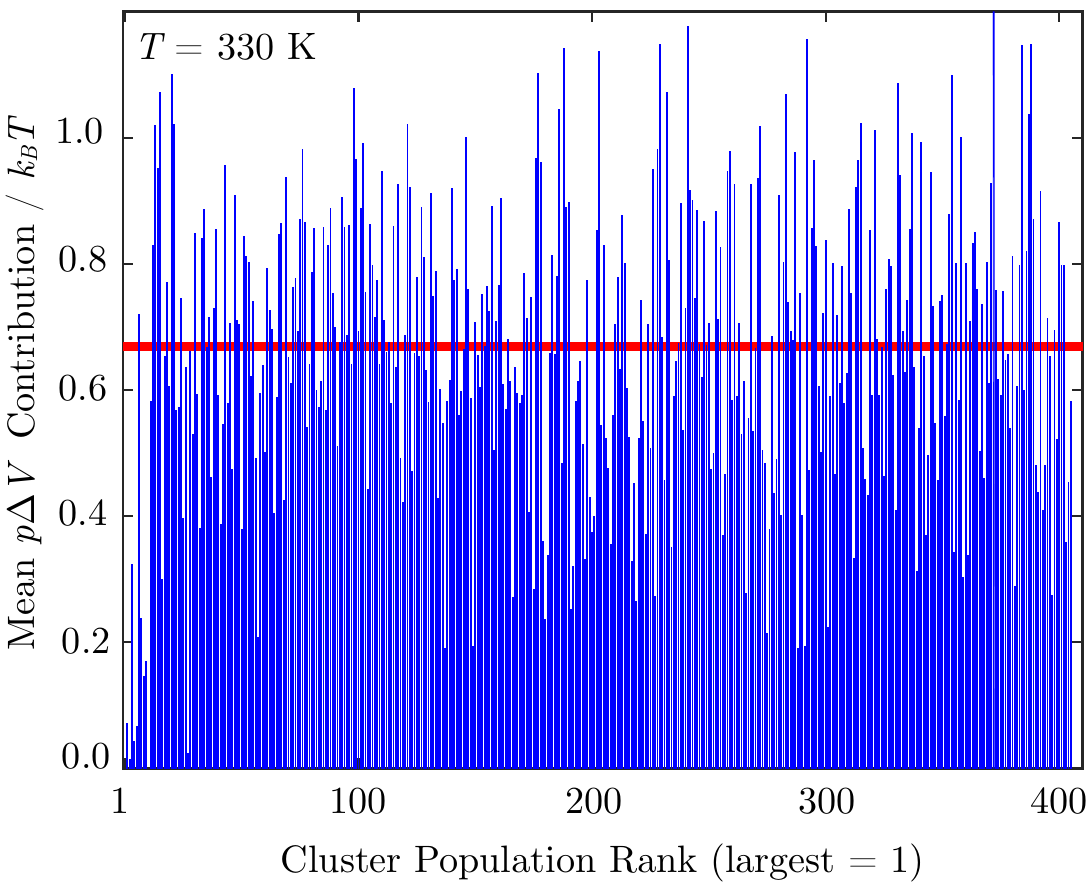}
\caption{ Magnitude of the $p\Delta V$ correction, calculated from REMD simulations at $T = 330$ K. This high temperature limit represents a worst--case scenario for cell volume fluctuations, where we expect large values to be attained for a $p \Delta V$ term in the spin--1 model.  The red line denotes the mean value, $\overline{p\Delta V}  = 0.669 k_B T$, from high--energy clusters ($\Delta G_j \geq 4\, k_BT$).}
\end{figure}